\documentclass[aps,twocolumn,floatfix,superscriptaddress]{revtex4-1}  
\usepackage{graphicx}                                                                                                            
\usepackage{amssymb,amsmath}                                                                                                     
\usepackage[colorlinks,citecolor=blue,urlcolor=blue]{hyperref}                                                                   
\usepackage{dcolumn}
\usepackage{color}       
\newcommand{\ve}[1]{\boldsymbol{#1}}

\begin{document}

\title{Local moments versus itinerant antiferromagnetism: magnetic phase diagram and spectral properties 
of the anisotropic square lattice Hubbard model}

\author{Marcin Raczkowski}
\affiliation{Institut f\"ur Theoretische Physik und Astrophysik,
             Universit\"at W\"urzburg, Am Hubland, D-97074 W\"urzburg, Germany}
\author{Fakher F. Assaad}
\affiliation{Institut f\"ur Theoretische Physik und Astrophysik and W\"urzburg-Dresden Cluster of Excellence ct.qmat,
             Universit\"at W\"urzburg, Am Hubland, D-97074 W\"urzburg, Germany}
\author{Masatoshi Imada}
\affiliation{Toyota Physical and Chemical Research Institute, 41-1 Yokomichi, Nagakute, Aichi, 480-1118, Japan}
\affiliation{Research Institute for Science and Engineering, Waseda University, 3-4-1 Okubo, Shinjuku-ku, Tokyo, 169-8555, Japan} 

\date{\today}

\begin{abstract} 
Using a cluster extension of the dynamical mean-field theory (CDMFT) we map out the magnetic phase diagram 
of the anisotropic square lattice Hubbard model with nearest-neighbor intrachain $t$ and interchain $t_{\perp}$ 
hopping amplitudes at half-filling. 
A fixed value of the next-nearest-neighbor hopping $t'=-t_{\perp}/2$ removes the nesting property of the Fermi surface 
and stabilizes a paramagnetic metal phase in the weak-coupling regime.   
In the isotropic and moderately anisotropic regions, a growing spin entropy in the metal phase is quenched out at 
a critical interaction strength by the onset of long-range antiferromagnetic (AF) order of preformed local moments. 
It gives rise to a first-order metal-insulator transition consistent with the Mott-Heisenberg picture.
In contrast, a strongly anisotropic regime $t_{\perp}/t\lesssim 0.3$  displays a quantum critical behavior 
related to the continuous transition between an AF metal phase and the AF insulator. 
Hence, within the present framework of CDMFT, the opening of the charge gap is magnetically driven as advocated 
in the Slater picture. We also discuss how the lattice-anisotropy-induced  evolution of the electronic structure on 
a metallic side of the phase diagram is tied to the emergence of quantum criticality.   
\end{abstract}

\maketitle

\section{Introduction}

The Hubbard model at half-filling and finite doping has been a subject of numerous theoretical studies 
driven by its ability to account for a variety of collective behaviors in strongly correlated quantum systems
such as a metal-insulator transition (MIT)~\cite{RMP98,Gebhard_book}, antiferromagnetism with its precursors and competitors~\cite{RMP18}, 
dimensional crossover from the one-dimensional (1D) Tomonaga-Luttinger liquid physics to a higher-dimensional situation~\cite{Giamarchi04},
pseudogap behavior in the single-particle spectral function~\cite{PRX18}, and high-$T_c$ superconductivity~\cite{RMP12}. 
Even though the MIT is simply a consequence of the energy competition between Coulomb interactions which tend to 
localize electrons and the kinetic energy term which favors electron itineracy, its actual form is 
situation dependent with microscopic details of the electronic structure playing a leading role~\cite{Kim18}.  
Accordingly, different mechanisms of the MIT  have been proposed to explain the formation of an insulating phase.  

In the Mott-Hubbard picture of a correlation-driven MIT, a strong on-site Coulomb repulsion $U$  splits the half-filled 
conduction band to open a gap in the electronic excitations between the resultant lower and upper Hubbard 
bands~\cite{Mott49,Hubbard63}. Thus, the Mott-Hubbard MIT  is driven solely by local electron correlations and does
not involve any spontaneous symmetry breaking.  Valuable insight into fundamental aspects of the Mott-Hubbard MIT has come from  
the dynamical mean-field theory (DMFT)~\cite{Georges96}.  
Although DMFT neglects nonlocal correlation effects and thus becomes exact only in the limit of 
infinite dimensions~\cite{Metzner89,Muller89}, it is able to capture the formation of local moments and 
resultant high-frequency features in the single-particle spectrum --- lower and upper Hubbard sidebands.
In addition, DMFT predicts the first-order MIT line due to the coexistence regime of the metallic and 
insulating solutions at low temperature. The first-order MIT line terminates at a critical end point $(T_c,U_c)$
similar to an ordinary liquid-gas transition. Consequently, one expects that the Mott critical end point belongs to 
the Ising universality class~\cite{Castellani79,Kotliar00,Papa08} with the double occupancy playing the role of a scalar order 
parameter of the transition. Subsequent studies based on quantum cluster techniques~\cite{Maier05} revealed that the inclusion 
of short-range correlations on top of the local dynamics does not change the order of the Mott-Hubbard MIT 
which remains first order~\cite{Park08,Balzer09}.

Another issue when it comes to address the microscopic description of the MIT comes from the spin degrees of freedom 
which typically mask the Mott-Hubbard MIT by driving a magnetic instability of the metallic phase. 
This is for example  known to happen in the Hubbard model on a two-dimensional (2D) square lattice at half-filling ---
due to the perfect nesting of the Fermi surface, the zero-temperature ground state displays long-range  (AF) order for an arbitrary 
small on-site repulsion $U$. Essentially this type of localization is accounted for already at the single-particle level 
within a Slater formalism~\cite{Slater51}:
the onset of the antiferromagnetism leads to a doubling of the unit cell size which opens up an exponentially small 
gap $\Delta\propto t\exp(-2\pi\sqrt{t/U})$ even for small but nonzero $U$  at the Fermi level~\cite{Hirsch85}. 
Increasing $U$ reduces the double occupancy  and the Slater antiferromagnet progressively evolves into a Mott-Heisenberg 
insulator~\cite{Gebhard_book}  with an AF gap of order of $U$ as verified within a variety of theoretical 
approaches~\cite{Hofstetter98,Moukouri01,Kyung03,Pruschke03,Korbel03,Borejsza03,Gull08,Varney09,Taranto12,Zhou14,Schafer15,Tocchio16,Seki18,Tanaka19,Congjun19}. 
In this regime, local moments form already on the temperature scale $T\sim U$ and thus the AF insulator at $T=0$ 
is accounted for by a spin-1/2  Heisenberg model of localized spins with the superexchange constant $J=4t^2/U$.

A prominent exception of the Mott insulator without any broken-symmetry-induced folding of the Brillouin zone down to $T=0$ 
is found in the 1D Hubbard model at half-filling~\cite{Lieb68}. 
The essential difference in nature between 1D and  higher-dimensional situations makes the studies of dimensional-crossover-driven phenomena 
very interesting~\cite{Metzner92,Boies95,Kohno07,Raczkowski13,Raczkowski15,Wohlfeld17,Dupont18,Ehlers18,Fm_af20}. 
Moreover, a combined effect of strong spatial anisotropy and geometrical frustration affects the strength of quantum fluctuations 
which in turn reduce the amplitude of broken-symmetry order parameters. This leaves a window for quantum critical behaviors and the 
emergence of novel ground states with anomalous single-particle spectra in their 
neighborhood~\cite{Biermann01,Essler02,Essler05,Berthod06,Tsuchiizu07,Penc11,Faye17,Feiguin19}. 
Our previous works in this domain~\cite{Raczkowski12,Lenz16} have established intriguing issues such as the unusual 
topology of the Fermi surface with dynamically generated pockets and fingerprints of Mott quantum criticality. 
Since the calculations in Refs.~\cite{Raczkowski12,Lenz16} were carried out in the paramagnetic (PM)  phase of the model, 
where by construction no long-range order is possible, it is conceivable that the onset of long-range order underlies the Mott transition. 
Thus, our aim is to revisit the phase diagram  by adapting a cluster extension of DMFT (CDMFT) 
to handle standard N\'eel-type AF order~\cite{Lichtenstein00}.

Strictly speaking, long-range magnetic order on a 2D lattice with short-range interactions is destroyed by long-wavelength fluctuations 
in the order parameter  at any finite temperature~\cite{MW66}. 
However, at sufficiently low temperature, $T_N^{\textrm{DMFT}}$,  captured relatively well already by a single-site DMFT~\cite{Rost12}, 
the correlation length grows exponentially with inverse temperature. 
This length scale defines an energy and time scale related to the fluctuations that are responsible for the destruction of 
the long-range order. Any experiment that is not able to resolve this energy or time scale will effectively perceive long-range order. 
Hence even at finite temperature, the fact that the correlation length diverges exponentially as a function of temperature 
has a very clear experimental signature observed in quantum simulations with ultracold atoms in optical 
lattices~\cite{Parsons16,Cheuk16,Mazurenko17,Drewes17}.

On the basis of CDMFT calculations, it was conceived in Ref.~\cite{Fratino17b} that the position of maximum of $T_N^{\textrm{DMFT}}(U)$ 
on the 2D square lattice is controlled by the critical end point $(T_c,U_c)$ of the MIT in the normal phase when the AF instability is
artificially suppressed. 
Assuming that this (in general hidden) Mott-Hubbard MIT indeed marks the qualitative change in the microscopic mechanism behind the stability 
of AF order, an intriguing question arises: can one expect a profound influence on the underlying physics of the AF phase when the 
critical end point $T_c$ of the  Mott-Hubbard MIT is driven down to zero upon increasing the degree of lattice anisotropy as in Ref.~\cite{Lenz16}? 
Most importantly, does this Mott quantum criticality affect the nature of the transition between a PM metal and 
the AF insulator? With this question in mind we proceed to discuss our findings.

The rest of the paper is organized as follows.
In Sec.~\ref{technical} we introduce the model, specify our implementation and technical details of the broken-spin-symmetry CDMFT 
algorithm, and define observables of interest. Our main results are discussed in Sec.~\ref{results}. 
We begin by presenting the anticipated finite-temperature phase diagram in the plane of interaction strength $U $ 
and hopping anisotropy $t_{\perp}/t$. 
Next, we elucidate the evolution of critical temperature and interaction strength $(T_c,U_c)$ terminating the first-order MIT 
upon varying the degree of lattice anisotropy.
Subsequently,  we turn our attention to the corresponding evolution of the electronic structure in the PM metal phase. 
Finally, we examine the reconstruction of low-energy quasiparticle excitations on going through the itinerant AF transition identified in the 
quasi-1D region of the phase diagram. 
We summarize our results and point out possible future directions in Sec.~\ref{discuss}.

\section{\label{technical} Model, method, and observables}

To handle numerically crucial physical ingredients at play we use here CDMFT. Specifically,  the $2\times2 $ cluster 
is a minimal unit cell which allows one to capture the 1D umklapp scattering process opening a gap 
in the half-filled band~\cite{Lieb68,Bolech03,Capone04,Go09}  and at the same time to treat short-range $x$- and $y$-axis 
AF spin fluctuations on an equal footing. 

Our aim is to  extend previous CDMFT studies of the influence of strong AF  correlations on the nature 
of the MIT in the 2D Hubbard model~\cite{Sato16,Fratino17a,Fratino17b} to the quasi-1D case.
To this end, we consider the square lattice  Hubbard model with an anisotropic hopping at half-filling, 
\begin{equation}
H=-\sum_{\pmb{ij},\sigma}t^{}_{\pmb{ij}}
   c^{\dag}_{{\pmb i}\sigma}c^{}_{{\pmb j}\sigma} +
   U\sum_{\pmb i}n^{}_{{\pmb i}\uparrow}n^{}_{{\pmb i}\downarrow} 
   -\mu\sum_{\pmb i,\sigma}n_{{\pmb i}\sigma},
\label{eq:Hubb}
\end{equation}
with a local Coulomb repulsion $U$, chemical potential $\mu$,  and electron hopping amplitudes: $t_{\pmb{ij}}=t$ on the intrachain bonds, 
$t_{\pmb{ij}}=t_{\perp}$ on the interchain bonds, and $t_{\pmb{ij}}=t'=-t_{\perp}/2$ between next-nearest-neighbor sites on 
two adjacent chains. Thus, the energy dispersion for the non-interacting case reads
\begin{equation}
	\epsilon_{\ve{k}} = -2t\cos k_x -2t_{\perp}\cos k_y - 4t'\cos k_x \cos k_y -\mu.	
\label{eq:eps}	
\end{equation}
A finite value of $t'$ breaks the perfect nesting property of the Fermi surface  $\epsilon_{\ve{k}}=- \epsilon_{\ve{k}+\ve{Q} }$.  
It also introduces a magnetically frustrating interaction. Both effects are expected to suppress the weak-coupling tendency 
towards the onset of low-temperature symmetry-broken states. Moreover, one could hope to find a region in the phase diagram 
where the impact of the next-nearest-neighbor hopping $t'$ is strong enough to push the magnetic transition temperature  
below the critical end point temperature $T_c$ thus exposing the Mott-Hubbard MIT~\cite{Zitzler04,Peters09}.
A downside of the lifted perfect nesting condition is that one has to adjust the chemical potential $\mu$ as a function of 
$t_{\perp}$, $U$, and $T$ to keep the required condition of a half-filled band. 

In CDMFT the original interacting lattice is mapped onto a cluster quantum impurity problem embedded in a self-consistent electronic bath. 
A predefined unit cell of volume given by the cluster size allows easily to study spin-symmetry-broken phases with a commensurate 
wavevector such as simple N\'eel AF order. 
To implement the CDMFT method, we  decompose the lattice into $N_u$ supercells with $N_c$ atoms each. As a result,  
the non-interacting Green's function ${\pmb G}_0 ({\pmb K},i \omega_m)$ and 
the spin-dependent self-energy ${\pmb \Sigma}_{\sigma} ( {\pmb K},i \omega_m)$ correspond to $N_c \times N_c $ matrices 
with wavevectors ${\pmb K}$ that span the reduced Brillouin zone of a supercell. In analogy to the DMFT approach, 
the CDMFT approximation neglects the  ${\pmb K}$ dependency of the self-energy, 
${\pmb \Sigma }_{\sigma}({\pmb K},i \omega_m) \equiv {\pmb \Sigma}_{\sigma}(i \omega_m) $.  
The latter is extracted by solving the effective cluster model: 
given the initial bath Green's function ${\pmb {\cal G}}_{0,\sigma} (i \omega_m)$,  we use a cluster impurity solver to  obtain 
the corresponding cluster Green's function  ${ \pmb {\cal G} }_{\sigma} (i \omega_m)$ and hence --- via the Dyson equation 
--- the cluster self-energy 
${\pmb \Sigma }_{\sigma} (i \omega_m) =  {\pmb {\cal G} }^{-1}_{0,\sigma}(i \omega_m) - {\pmb {\cal G} }^{-1}_{\sigma}(i \omega_m)$. 
The self-consistent loop is closed by requiring that the cluster Green's function ${ \pmb {\cal G} }_{\sigma} (i \omega_m)$ 
matches the lattice Green's function of the  original model formulated in the cluster-site basis: 
\begin{align}
\label{loop}
	{ \pmb {\cal G} }_{\sigma} (i \omega_m) & = 
	 \frac{1} { { \pmb {\cal G} }^{-1}_{0,\sigma}(i \omega_m) - { \pmb \Sigma_{\sigma}(i \omega_m)} } \nonumber \\
     & = \frac{1}{N_u} \sum_{\pmb{K}} 	  
	 \frac{1} { {\pmb G}_0^{-1}(\pmb{K},i \omega_m) - {\pmb \Sigma}_{\sigma} (i \omega_m) }.
\end{align} 
This allows us to compute a new bath Green's function ${\pmb {\cal G}}_{0,\sigma} (i \omega_m)$ 
which is fed back to the impurity solver and the   procedure is repeated till convergence is reached.  

While our preliminary results for the 2D case were obtained using the quantum Monte Carlo (QMC) method of Hirsch and Fye as cluster 
solver~\cite{HF86}, we found it advantageous to switch to the continuous-time QMC  (CT-QMC) algorithm~\cite{RMP11}. 
In particular we opted for its weak-coupling implementation based on a stochastic series expansion for 
the partition function in the interaction representation~\cite{Rubtsov05,Assaad07}.  
It allowed us to reach temperatures as low as $T=t/100$ necessary to pin down the evolution of $T_c$ 
in the quasi-1D region.  In addition, the CT-QMC algorithm provides the possibility of Monte Carlo measurements directly 
on the Matsubara-frequency $\omega_m$ axis. Thus  one avoids the cumbersome transformation from imaginary time to  Matsubara frequencies 
necessary when the Hirsch-Fye solver is used instead.  
Note that the next-nearest-neighbor hopping $t'$ in Eq.~(\ref{eq:eps}) breaks the particle-hole symmetry of the Hamiltonian and thus 
introduces the negative sign problem in the QMC simulations. Results of the average sign in our studies are shown in Appendix~\ref{app:sign}.

To determine the domain of stability of the AF phase, in the actual simulation we explicitly break the SU(2) spin symmetry  
by introducing a small staggered magnetic field through the initial guess for the bath Green's function ${\pmb {\cal G}}_{0,\sigma} (i \omega_m)$. 
Hence, if the parameters of a simulation correspond to the regime with a thermodynamically stable AF phase,  
in the course of the CDMFT iterative process one converges to the solution with a finite staggered magnetization
\begin{equation}
m=\frac{1}{N_c}\sum_{\pmb{i}} (-1)^{\pmb i}
 \langle n^{}_{{\pmb i}\uparrow} - n^{}_{{\pmb i}\downarrow}\rangle.
\end{equation}
We remark that allowing for a broken-spin-symmetry electronic bath substantially simplifies studies of otherwise 
a very intricate --- in a generic nonrelativistic case --- problem  of a metal at the spin-density-wave quantum 
critical point involving coupling of gapless long-wavelength Goldstone modes 
to a Fermi sea~\cite{Metlitski10,Hartnoll11,Berg12,Lee13,Gerlach17,Schlief17}.

Further insight into the underlying physics in different parts of the diagram is obtained from the following observables:

(i) Double occupancy
\begin{equation} 
D=\frac{1}{N_c}\sum_{\pmb{i}}\langle n^{}_{{\pmb i}\uparrow}n^{}_{{\pmb i}\downarrow}\rangle.
\end{equation}

(ii) Lattice Green's functions $g_{\sigma}({\pmb k}, i \omega_m )$ in the original Brillouin zone 
with  ${\pmb k} \in \left[ - \pi,\pi \right]$. For  consistency with our previous studies~\cite{Raczkowski12,Lenz16}, 
we extract it by periodizing the Green's function in the cluster-site basis:  
\begin{align}
   g_{\sigma}({\pmb k}, i \omega_m )                 &=  
\frac{1}{N_c} \sum_{\mu,\nu = 1}^{N_c} e^{i {\pmb k} 
        \left( {\pmb a}_\mu - {\pmb a}_\nu\right) }  \nonumber \\ 
                                                     &\times 
      \left[\frac{1} { {\pmb G}_0^{-1}( {\pmb K} , i \omega_m) - 
       {\pmb \Sigma}_{\sigma} ( i \omega_m) } \right]_{\mu,\nu},
\end{align} 
where ${\pmb a}_\mu,{\pmb a}_\nu$ label cluster sites.  It is however fair to remark that there are other 
periodization schemes such as the cumulant periodization which yield a faster convergence against the cluster size 
to the thermodynamic limit, an issue that becomes crucial in the proper description of doped 2D Mott insulators~\cite{Stanescu06a,Sakai12}. 
Given that we are interested here in spectral properties of a moderately correlated metallic phase stabilized by strong magnetic 
frustration, one might still hope that the applied Green's function periodization scheme reproduces the qualitative features of 
the thermodynamic solution.  

(iii) Momentum-resolved spectral function at the Fermi level $A_{\pmb k}(\omega=0)$. 
We estimate it from the behavior of lattice Green's function at large imaginary time $\tau$ which 
gives the integrated spectral intensity in a frequency window of width $~T$ around the Fermi level~\cite{GB2} 
$A_{\pmb k}(\omega=0) \propto \lim\limits_{\beta\to\infty}\beta g(\pmb{k},\tau=\beta/2 ) $ where $\beta=1/T$ and 
\begin{equation}
	g(\pmb{k},\tau=\beta/2) = \frac{1}{\beta}\sum_{\omega_m,\sigma } e^{-i\omega_m(\tau=\beta/2)} g_{\sigma}({\pmb k}, i \omega_m ).
\end{equation}
We use  $g(\pmb{k},\tau=\beta/2)$  to analyze the evolution of the Fermi surface across the phase diagram.  

(iv) Density of states at the Fermi level 
$N(\omega=0)=\frac{1}{N}\sum_{\pmb{k}} A_{\pmb k}(\omega=0)  \propto \lim\limits_{\beta\to\infty}\beta g_0(\tau=\beta/2 )$ where 
\begin{equation}
g_0(\tau=\beta/2) = \frac{1}{\beta N}\sum_{{\pmb k},\omega_m,\sigma } e^{-i\omega_m(\tau=\beta/2)} g_{\sigma}({\pmb k}, i \omega_m ).
\end{equation}

(v)  Momentum-resolved spectral function $A({\pmb k},\omega) = - \tfrac{1}{\pi} {\rm Im} g({\pmb k}, \omega )$. 
We have used the stochastic MaxEnt implementation~\cite{Sandvik98,Beach04a} of the Algorithms for Lattice Fermions (ALF) 
library~\cite{ALF2017} to rotate the imaginary-time Green's function  $g(\pmb{k},\tau)$  to the real-frequency axis.

\section{\label{results} Numerical results}

\subsection{\label{diagram} Magnetic phase diagram}

\begin{figure}[t!]
\begin{center}
\includegraphics[width=0.4\textwidth]{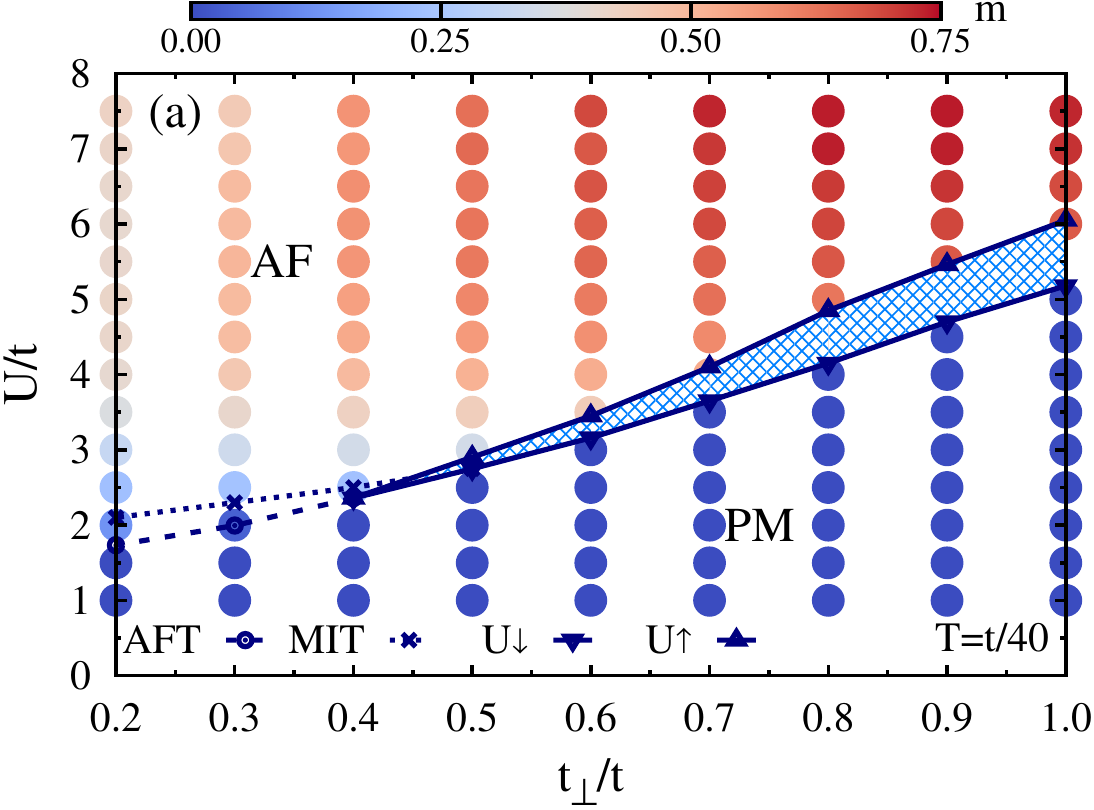}\\
\includegraphics[width=0.4\textwidth]{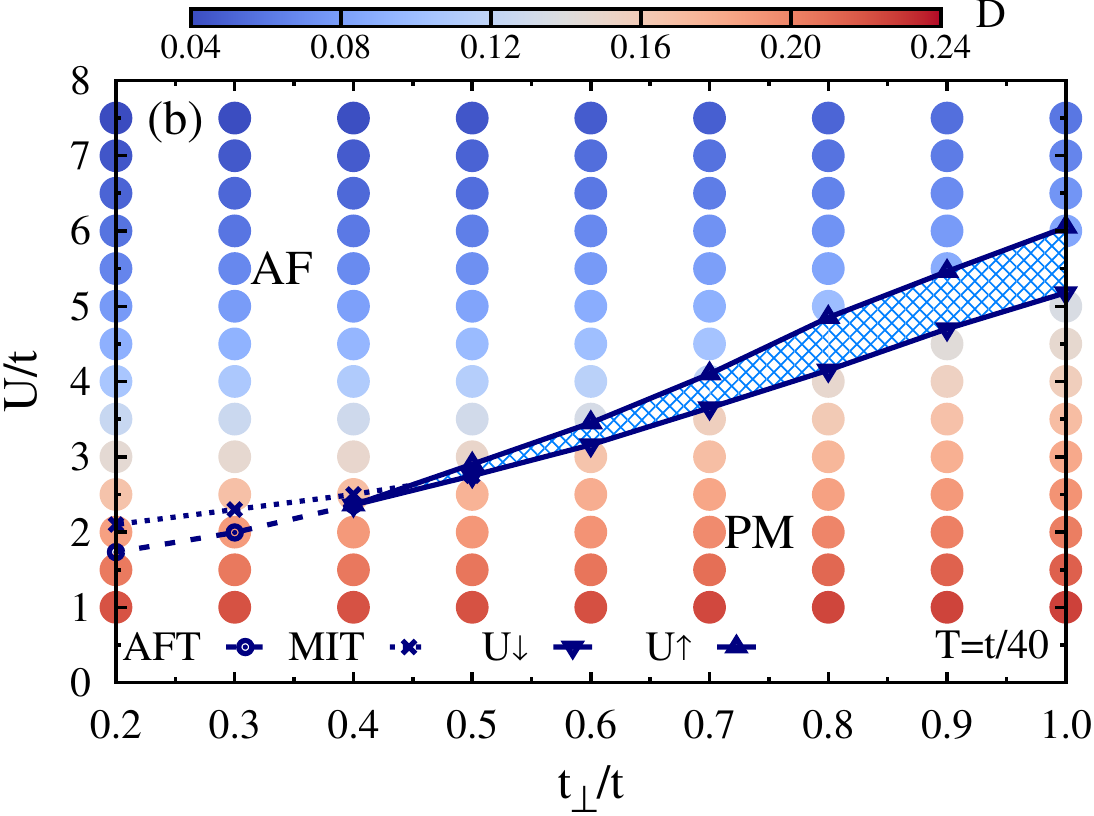}\\
\includegraphics[width=0.4\textwidth]{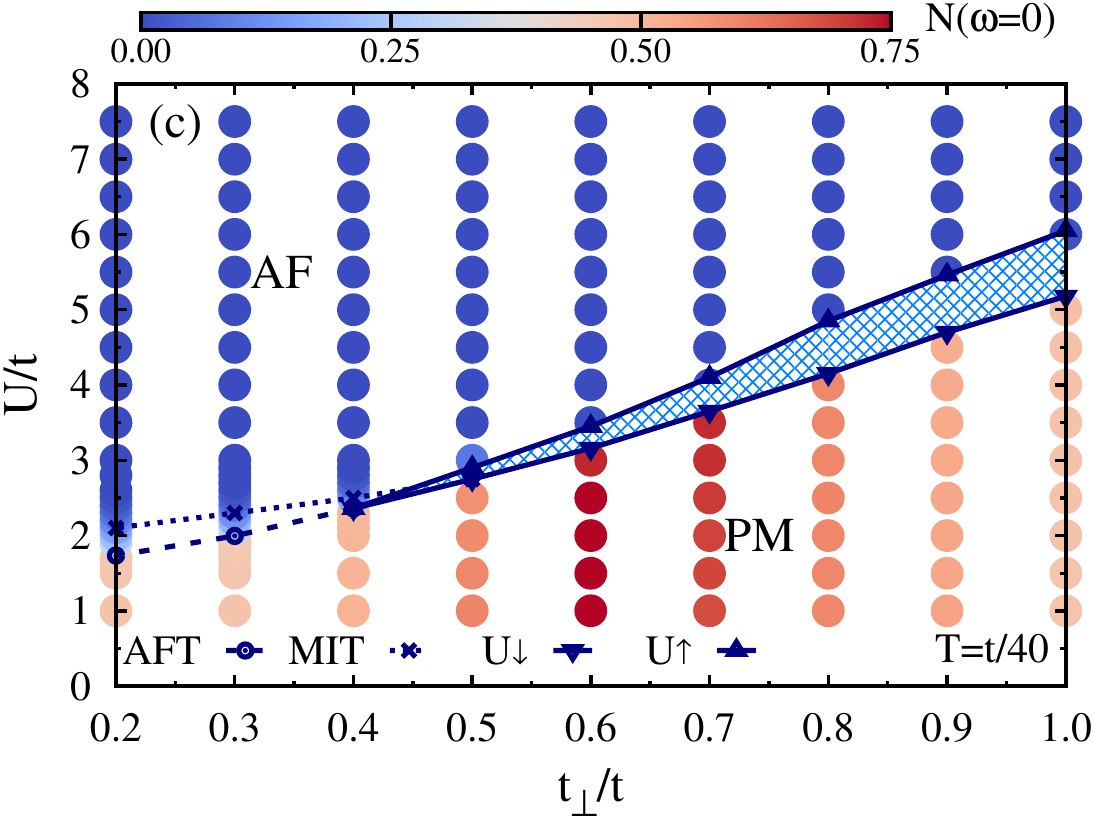}
\end{center}
\caption
{Finite-temperature CDMFT phase diagram at $T=t/40$ of the anisotropic 2D Hubbard model Eq.~(\ref{eq:Hubb}) at half-filling 
encompassing paramagnetic (PM) and antiferromagnetic (AF) phases. 
The blue shaded area between the ${U\hspace{-0.3em}\uparrow}$  (${U\hspace{-0.3em}\downarrow}$) lines corresponding 
to a simulation with increasing (decreasing) the interaction strength $U$ (see Appendix~\ref{app:hysteretic} for illustrative raw data), 
respectively, shows a coexistence region and implies a first-order metal-insulator transition (MIT). 
For interchain couplings $t_{\perp}/t\lesssim 0.4$, a continuous AF transition (AFT) (dashed line) enables the onset of an AF metal 
phase which eventually undergoes a second-order MIT (dotted line) to the AF insulator. 
Color-coded circles display the behavior of: 
(a) staggered magnetization $m$; 
(b) double occupancy $D$, and   
(c) density of states at the Fermi level $N(\omega=0)\propto \lim\limits_{\beta\to\infty}\beta g_0(\tau=\beta/2 )$.
}
\label{PD_cdmft}
\end{figure}

\begin{figure*}[t!]
\begin{center}
\includegraphics[width=0.32\textwidth]{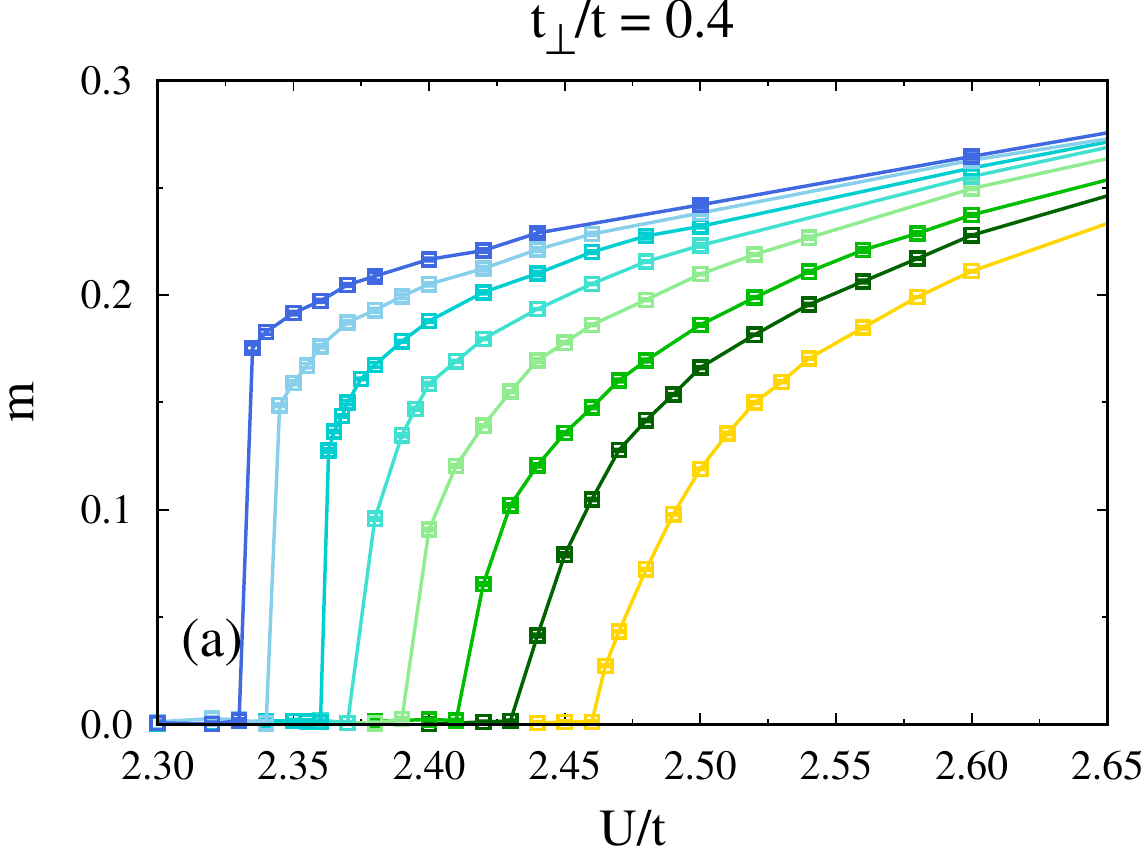}
\includegraphics[width=0.32\textwidth]{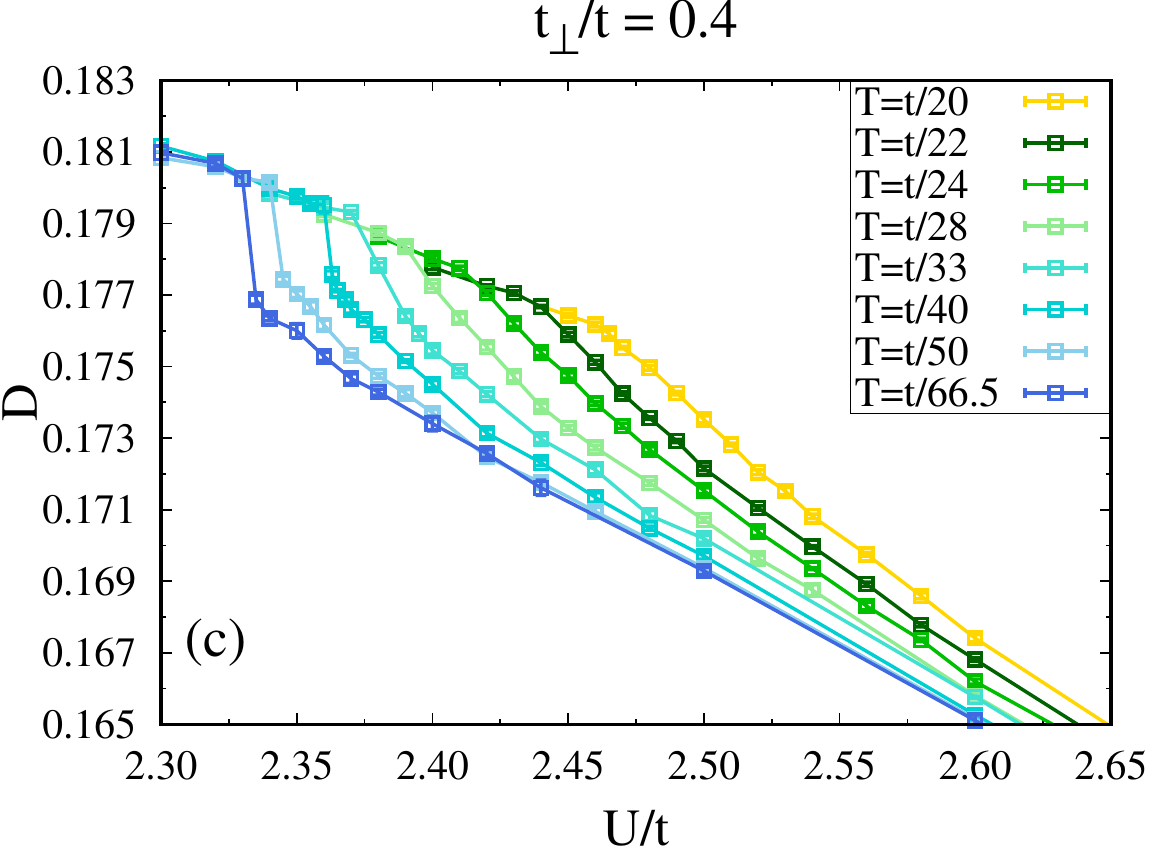}
\includegraphics[width=0.32\textwidth]{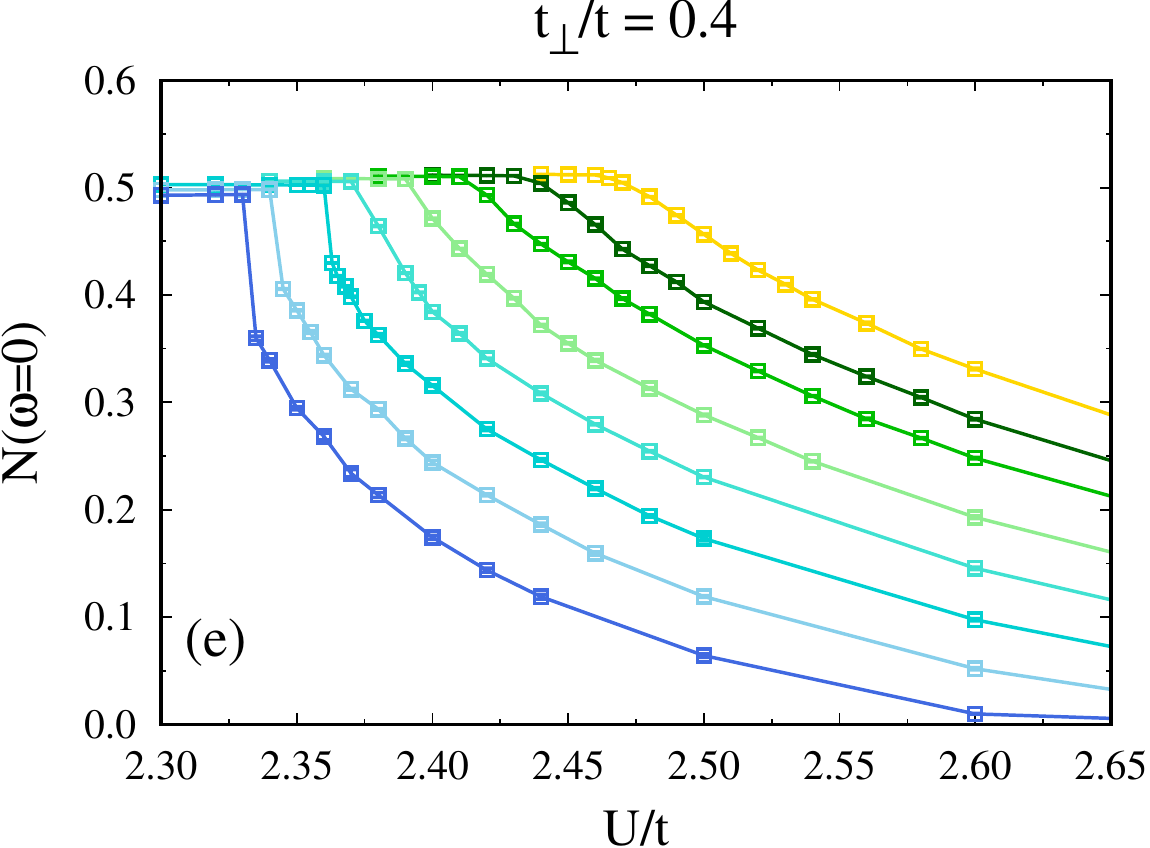}\\
\includegraphics[width=0.32\textwidth]{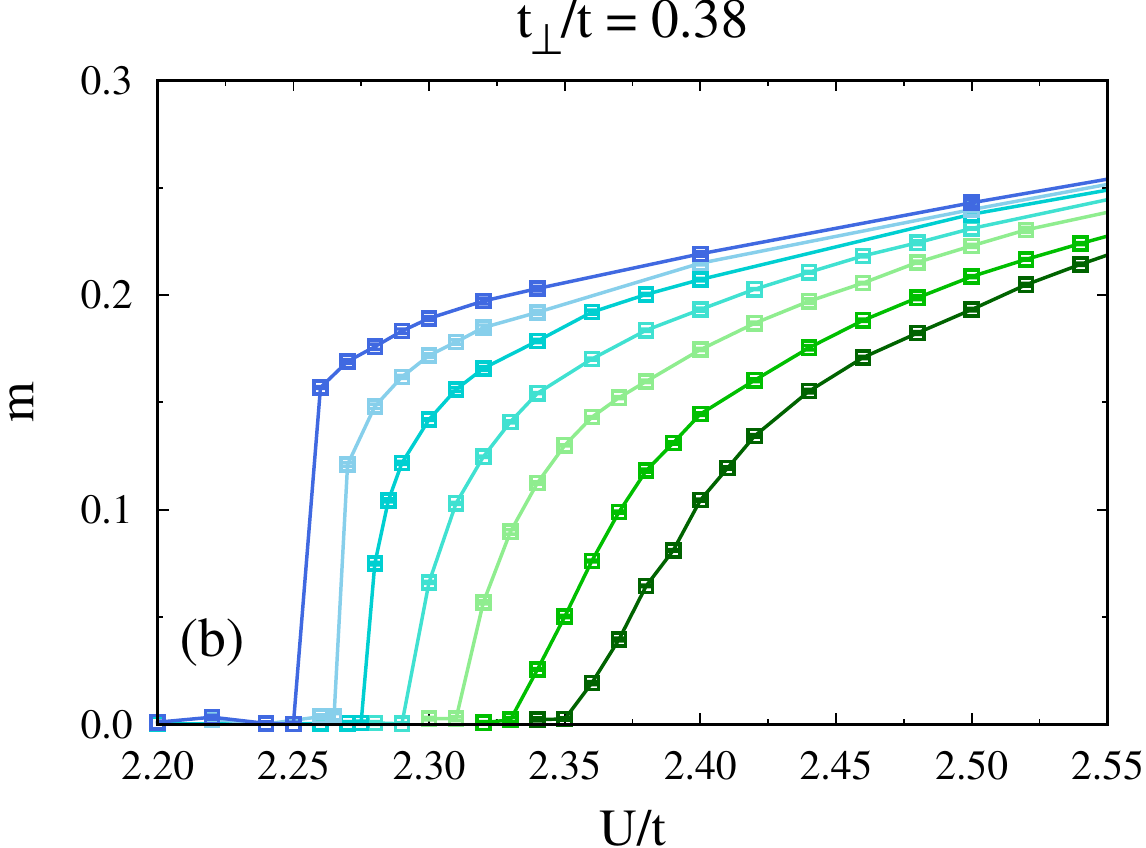}
\includegraphics[width=0.32\textwidth]{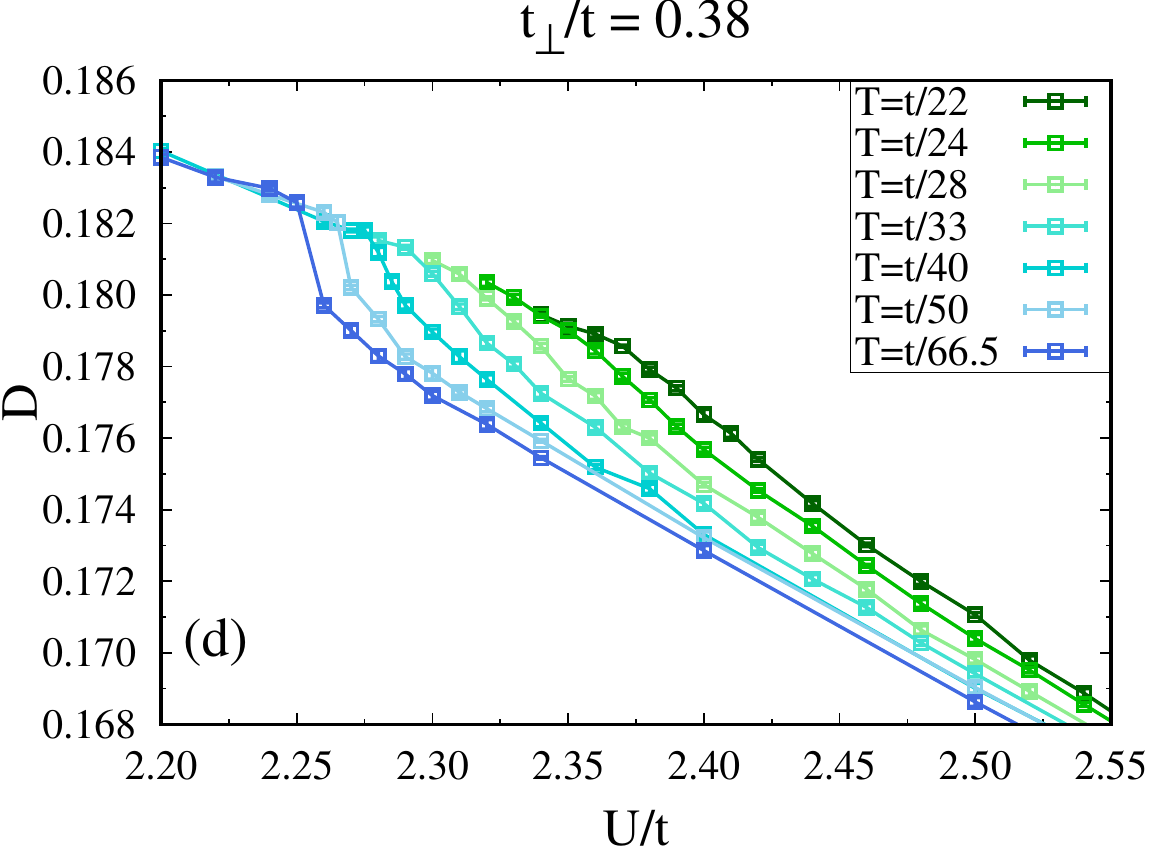}
\includegraphics[width=0.32\textwidth]{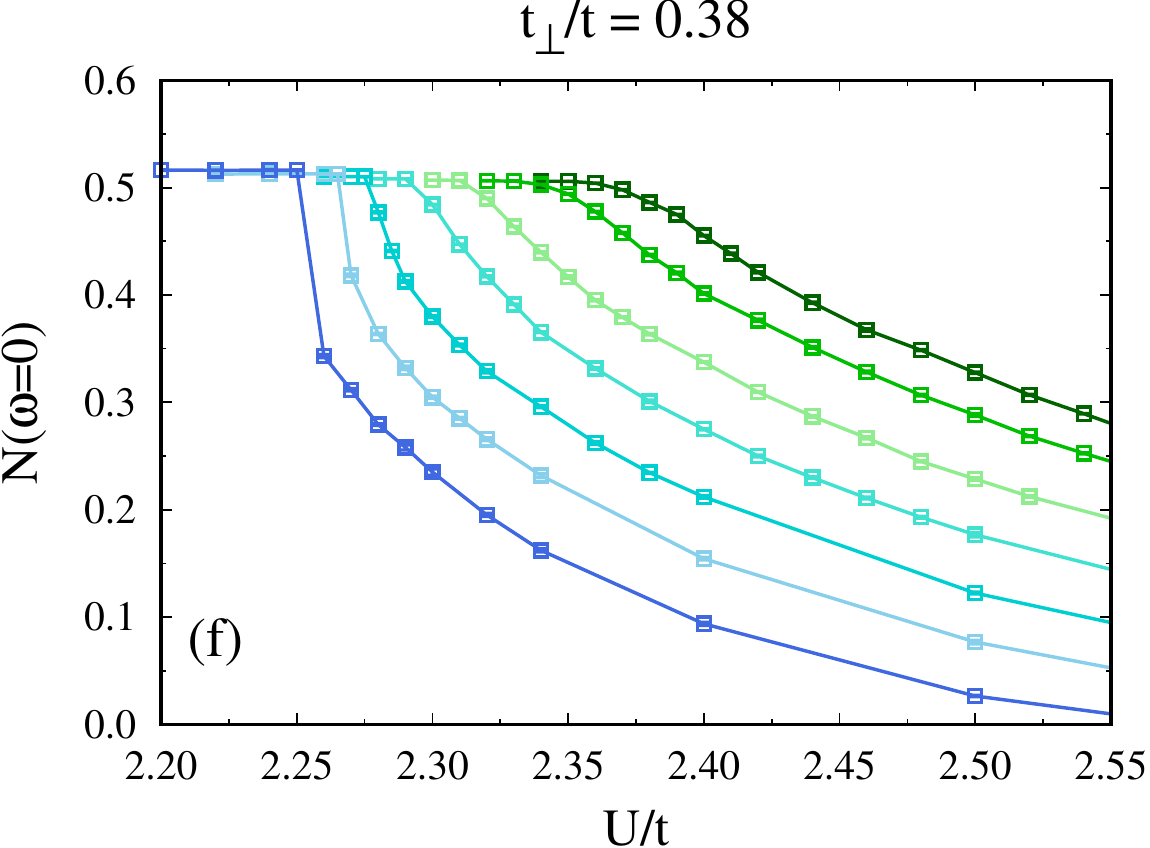}
\end{center}
\caption
{ 
Temperature dependence of the:
(a) and (b) staggered magnetization $m$;
(c) and (d) double occupancy $D$, and
(e) and (f) density of states at the Fermi level $N(\omega=0)\propto \lim\limits_{\beta\to\infty}\beta g_0(\tau=\beta/2 )$
obtained on decreasing $U$  for $t_{\perp}/t=0.4$ (top) and  $t_{\perp}/t=0.38$ (bottom) in the proximity to $T_c\simeq t/40$. 
}
\label{mag04}
\end{figure*}

Our CDMFT results are summarized in the $(U,t_{\perp})$ phase diagram shown in Fig.~\ref{PD_cdmft}. 
It was mapped out at a fixed temperature $T=t/40$.  On the one hand, a strong magnetic frustration introduced by the next-nearest-neighbor 
hopping $t'=-t_{\perp}/2$ stabilizes in the weak-coupling region a PM metal phase which as shown in Fig.~\ref{PD_cdmft}(a) extends 
to a fairly large interaction $U_c/t=5.17$ in the 2D limit. In this case, one finds a strong reduction of the double occupancy 
$D$ at $U_c$ indicative of the local moment $\langle S_z^2\rangle=1-2D$ formation, see Fig.~\ref{PD_cdmft}(b).
On the other hand,  we find that the Mott-Hubbard transition is preempted by AF order.  
Indeed, the $U\hspace{-0.3em}\uparrow$  line in Fig.~\ref{PD_cdmft} corresponds to the 
transition at a given $t_{\perp}$  from a PM metal phase to the AF insulator  with increasing $U$ while the $U\hspace{-0.3em}\downarrow$ 
line to the transition from the AF insulator to the PM metal phase  with decreasing $U$. 
Examples of such a hysteresis cycle observed in the raw data of the staggered magnetization $m$ and 
double occupancy $D$ are shown in Appendix~\ref{app:hysteretic}. Collecting the results for different values of $t_{\perp}$ 
lead us to the hysteretic region indicated as the blue shaded area in Fig.~\ref{PD_cdmft}. 
This hysteresis and jumps in both $m$ [Fig.~\ref{PD_cdmft}(a)]  and $D$ [Fig.~\ref{PD_cdmft}(b)] 
resolved for the moderately anisotropic region  $0.5\leq t_{\perp}/t\leq 1$ clearly imply a first-order AF transition 
concurrent with the MIT. This contrasts with a static mean-field approximation where depending on the specific form of the 
band structure tuned by the magnitude of $t'$, both the first order and continuous transition between a PM metal and the AF insulator 
can be reproduced~\cite{Kondo96,Hofstetter98}. In particular, within Hartree-Fock theory of the isotropic 2D Hubbard model 
one finds only a continuous transition for the particular choice $t'=-t/2$ used here. 
In Sec.~\ref{iso} we provide evidence that the first-order character of the MIT is actually a consequence 
of dominant local temporal fluctuations going beyond the mean-field approximation.

The situation in the strongly anisotropic part of the phase diagram is more subtle and requires more attention. 
In particular, a weak staggered magnetization in the vicinity of the AF transition (AFT) (dashed line in Fig.~\ref{PD_cdmft})  
accompanied by a relatively large double occupancy on the AF side are both suggestive of the itinerant magnetism.   

In order to identify the character of the transition for $t_{\perp}/t=0.4$, we examine in Fig.~\ref{mag04} 
the behavior of the staggered magnetization $m$ [Fig.~\ref{mag04}(a)], double occupancy $D$ [Fig.~\ref{mag04}(c)],  
and density of states at the Fermi level $N(\omega=0)$ [Fig.~\ref{mag04}(e)] upon decreasing $U$ for various temperatures. 
One finds that the smooth increase of $m$ at the highest $T=t/20$ gets steeper at lower temperatures and is replaced by 
a small discontinuity at $T=t/40$. The latter is accompanied by a jump seen both in $D$ and $N(\omega=0)$. 
This implies a first-order phase transition with a critical end point $T_c\simeq 1/40$ even though 
a slow convergence of the CDMFT self-consistency loop makes it hard to firmly assess the existence and range of the hysteresis 
at temperatures a little bit below $T_c$.     

In contrast, simulations with a slightly smaller $t_{\perp}/t=0.38$ yield a continuous onset of magnetism at $T=t/40$ 
and lower $T=t/50$ is required to resolve the discontinuity in all the three observables, 
see Figs.~\ref{mag04}(b), \ref{mag04}(d), and \ref{mag04}(f).   
This result together with a shrinkage of the hysteretic region seen in Fig.~\ref{PD_cdmft} is suggestive of a systematic reduction 
of critical end point $T_c$ as a function of the growing lattice anisotropy. We analyze this issue in more detail in Sec.~\ref{T_c}.

A continuous nature of the AF transition identified in the strongly anisotropic part of the phase diagram  
paves the way to an intermediate AF metal phase. Indeed, as shown in Fig.~\ref{PD_cdmft}(c) there is a narrow region 
in the vicinity of the AF transition with small but finite $N(\omega=0)$.
Upon further increase of the interaction strength, a crossover from the AF metal to the AF insulator occurs  
once the staggered magnetic moment is sufficient to fully gap out hole and electron Fermi pockets of the AF metal phase.   

Thus, in analogy with earlier studies restricted to PM solutions~\cite{Raczkowski12,Lenz16}, 
the continuous nature of the MIT stems from a smooth vanishing of the volume of Fermi pockets at the critical interaction. 
However,  unlike in Ref.~\cite{Lenz16}, where a dynamically generated breakup of the Fermi surface was the consequence of 
remnant 1D umklapp scattering, electron and hole Fermi pockets form here due to the doubling of the unit cell 
driven by  AF order.  As we discuss in Sec.~\ref{aniso}, it results in a different topology of the Fermi surface in the vicinity of 
the MIT.

\begin{figure*}[t!]
\begin{center}
\includegraphics[width=0.32\textwidth]{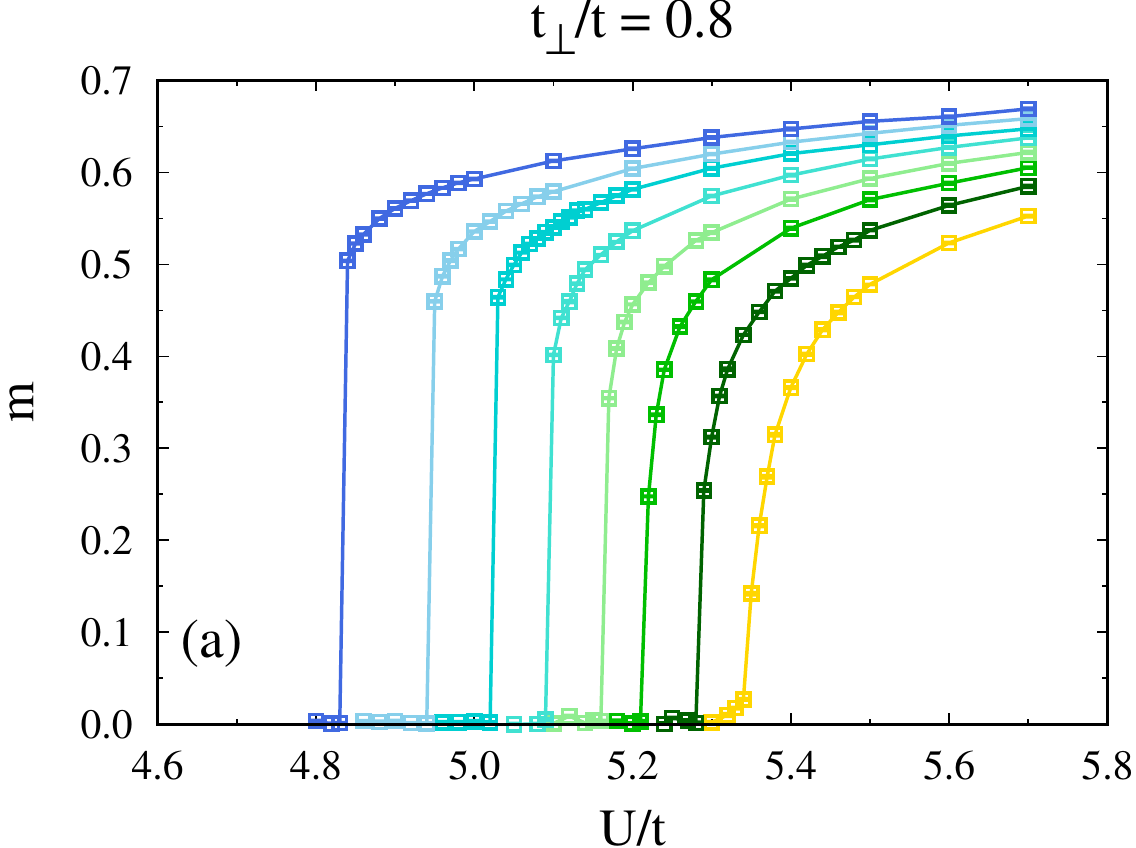}
\includegraphics[width=0.32\textwidth]{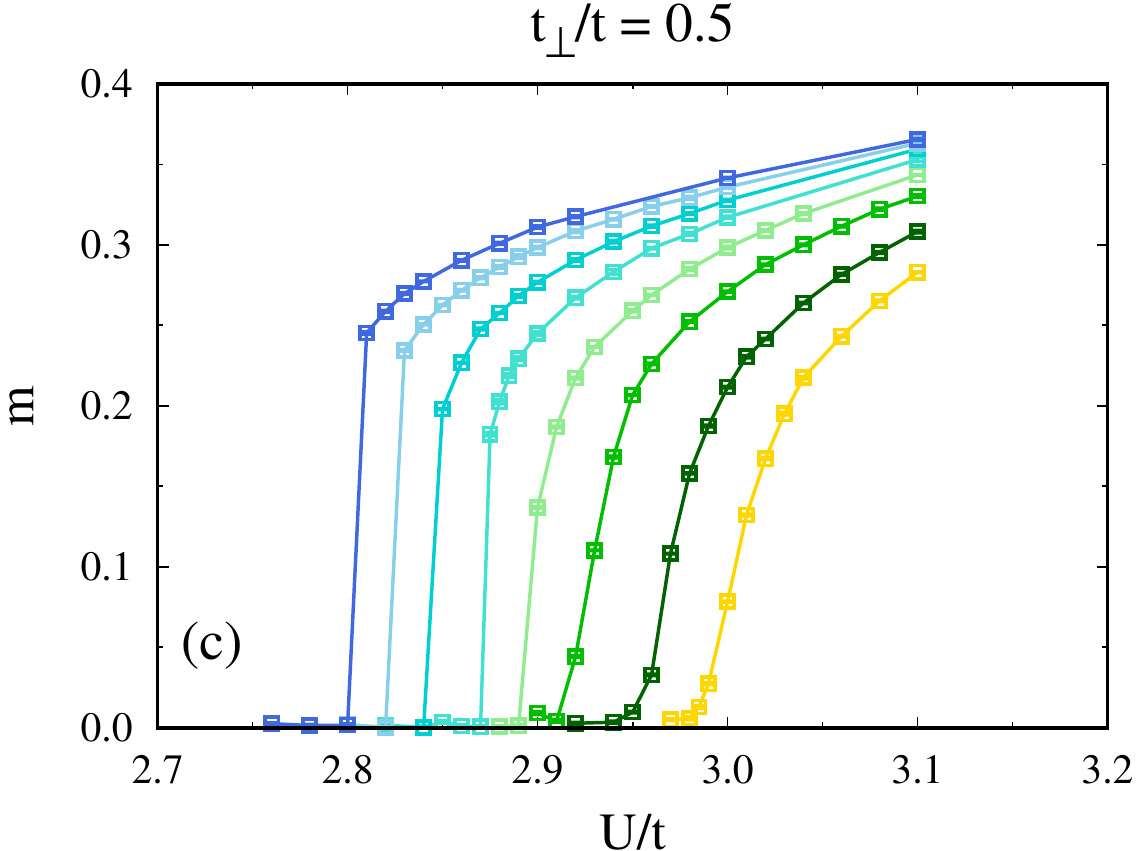}
\includegraphics[width=0.32\textwidth]{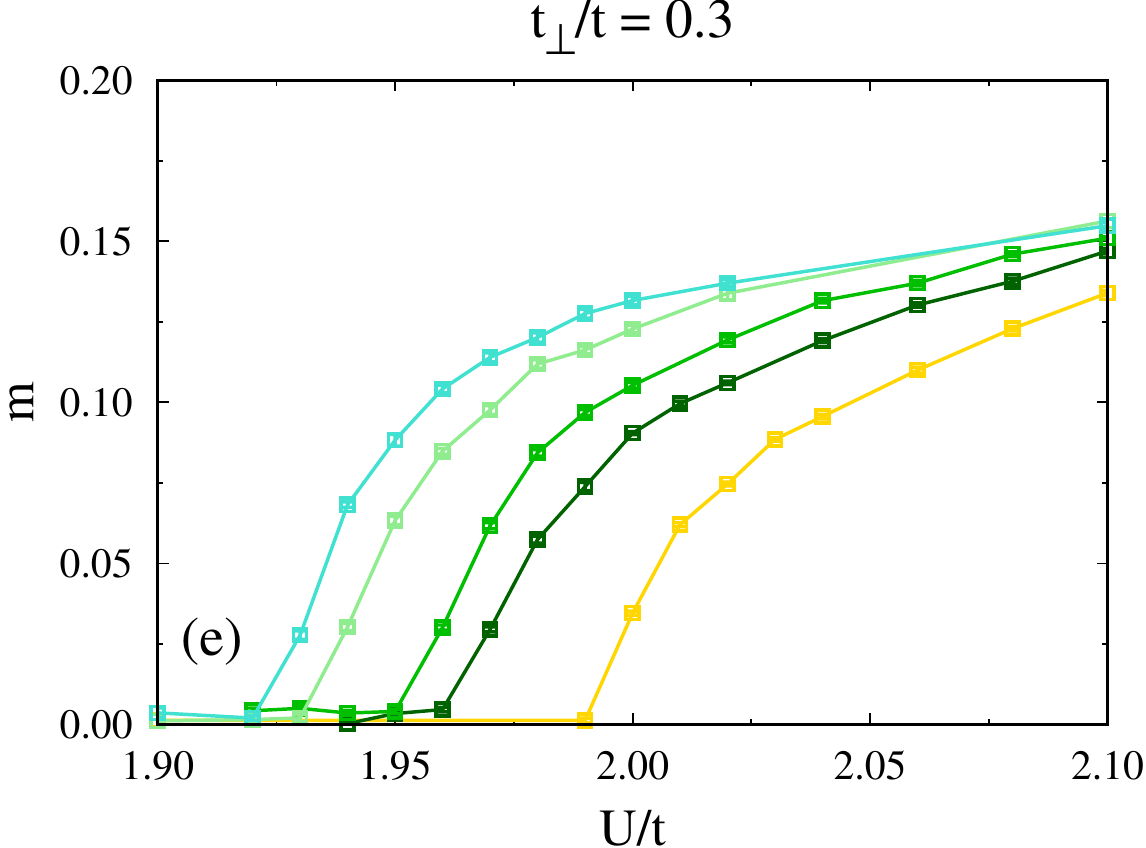} \\
\includegraphics[width=0.32\textwidth]{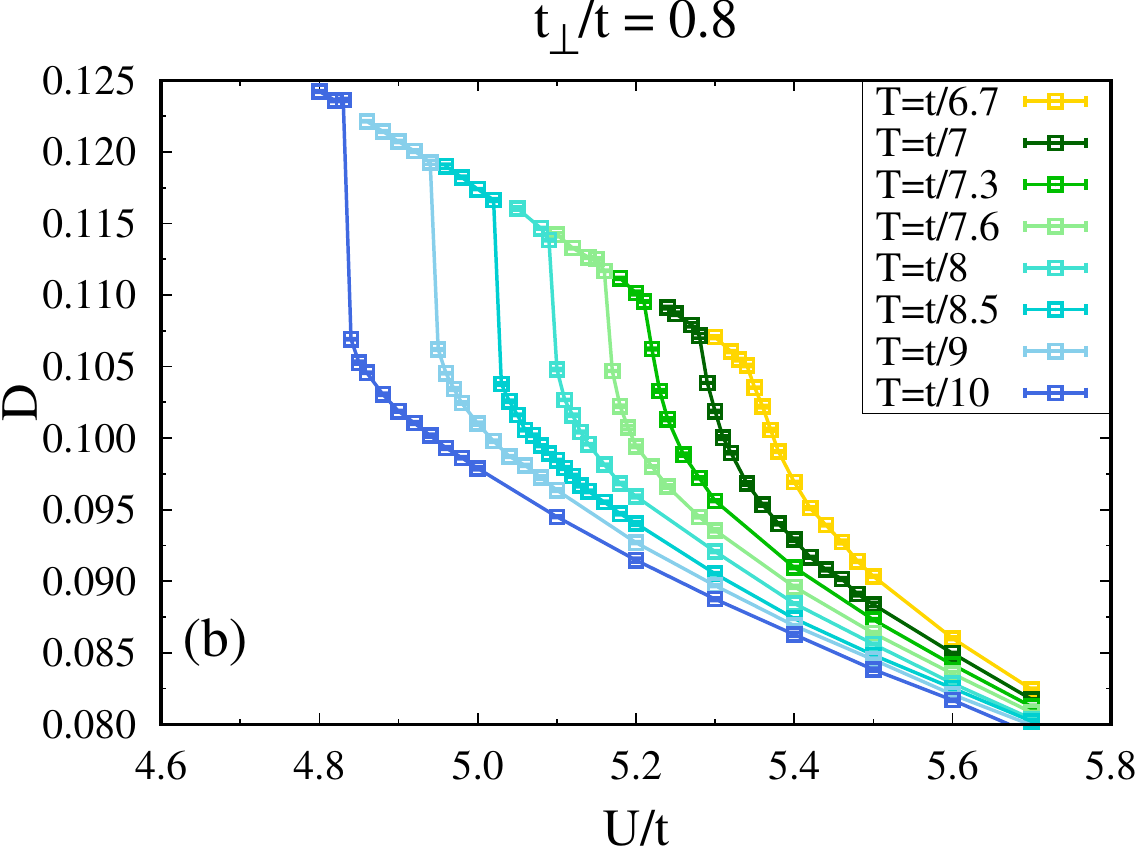}
\includegraphics[width=0.32\textwidth]{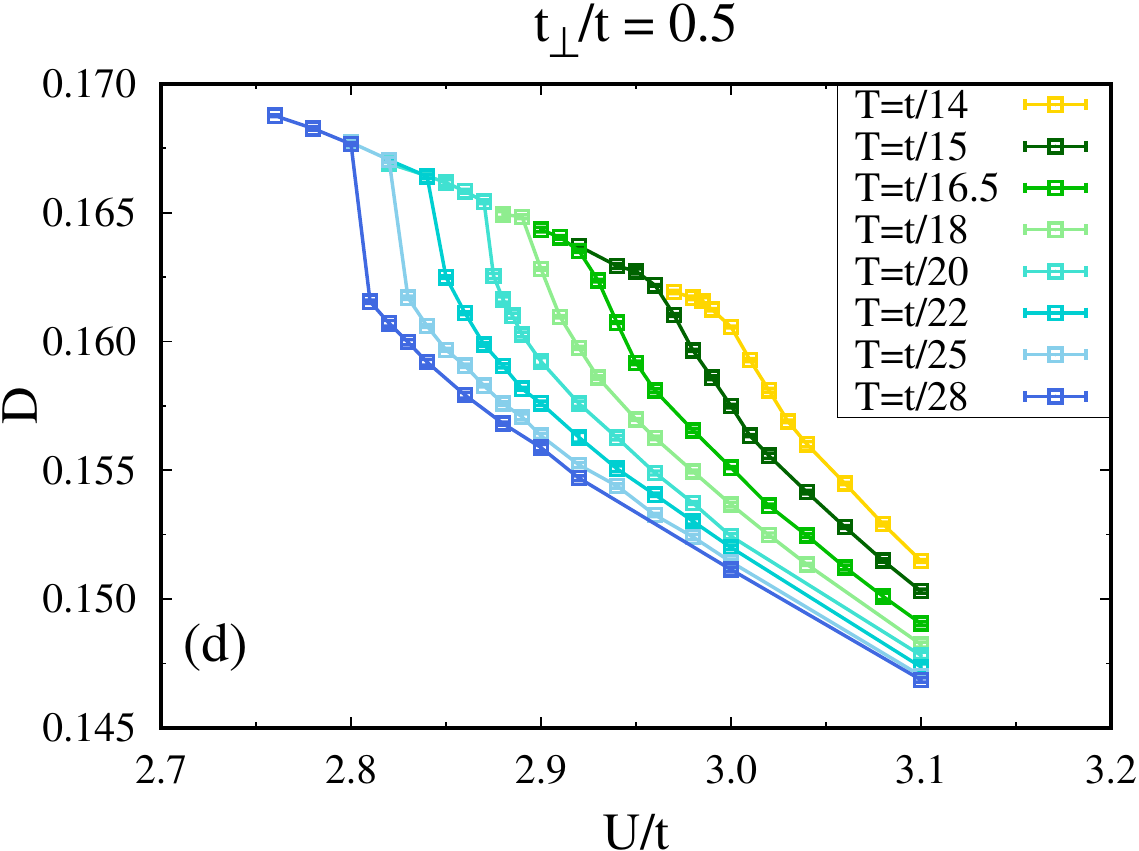}
\includegraphics[width=0.32\textwidth]{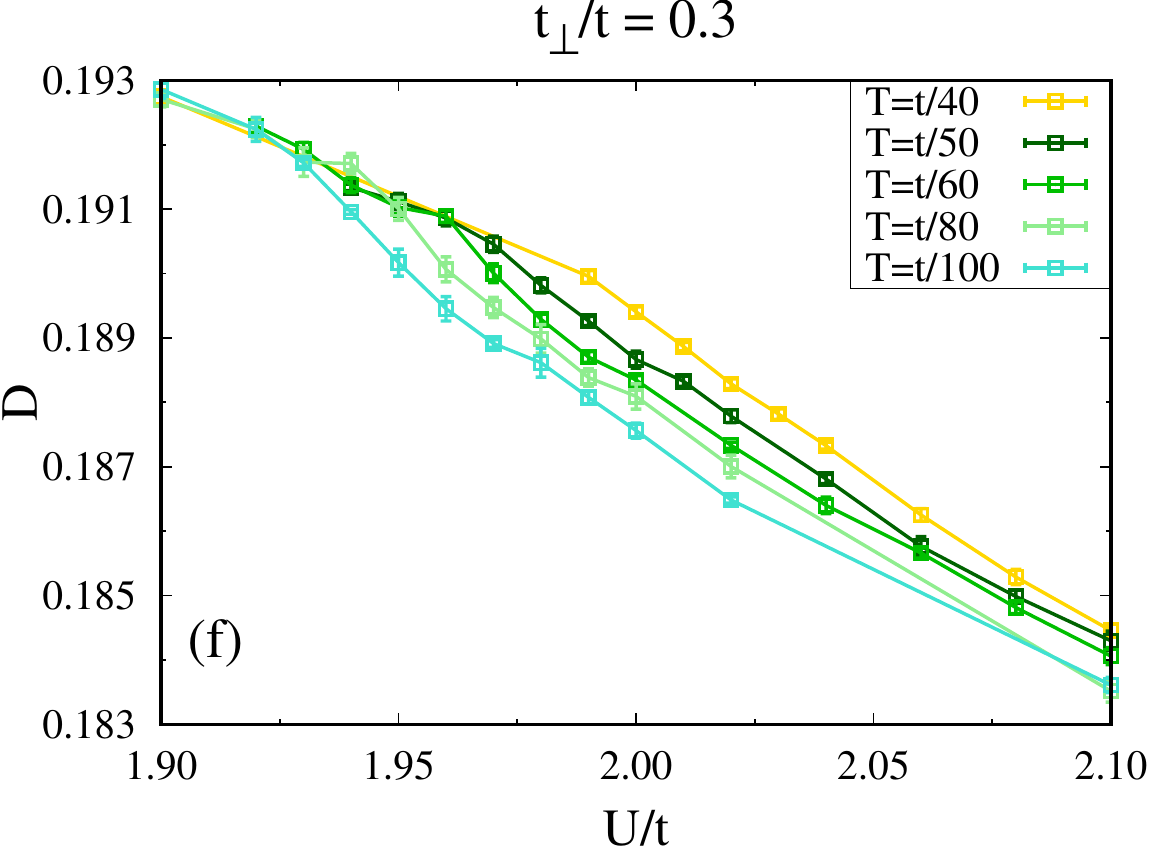}
\end{center} 
\caption
{Temperature dependence of the staggered magnetization $m$ (top) and  double occupancy $D$  (bottom)
obtained on decreasing $U$ in the case of: 
(a) and (b) weak anisotropy $t_{\perp}/t=0.8$; 
(c) and (d) moderate anisotropy $t_{\perp}/t=0.5$, and  
(e) and (f) strong anisotropy $t_{\perp}/t=0.3$.
}
\label{mag0853}
\end{figure*}

Finally, let us point out another important consequence of varying the ratio between inter- and intrachain hopping amplitudes 
--- the  existence of a  certain critical point where the non-interacting 2D closed Fermi surface undergoes a topological change 
into an open surface. 
For the specific choice $t'=-t_{\perp}/2$ used in our studies, this is known to happen at $t_{\perp}/t\simeq 0.62$~\cite{Ehlers18}.   
As we demonstrate in Sec.~\ref{moderate}, this topological (Lifshitz) transition is accompanied by a van Hove singularity in the single-particle 
density of states crossing the Fermi level. From a weak-coupling point of view,  a large  density of states  might lead to divergent 
non-interacting susceptibilities in both particle-hole and particle-particle channels signaling enhanced ordering tendencies. 
Nevertheless, w find that phase diagram boundaries are insensitive to the van Hove singularity passing smoothly across the region  
$0.6\le t_{\perp}/t\le 0.7$ with enhanced $N(\omega=0)$, see Fig.~\ref{PD_cdmft}(c). Together with a reduced double occupancy, 
this is yet another indication that the AF instability  in this part of the phase diagram is not of a weak-coupling origin but instead 
it should be considered as driven by the ordering tendency of preformed local moments.

\begin{figure}[t!]
\begin{center}
\includegraphics[width=0.45\textwidth]{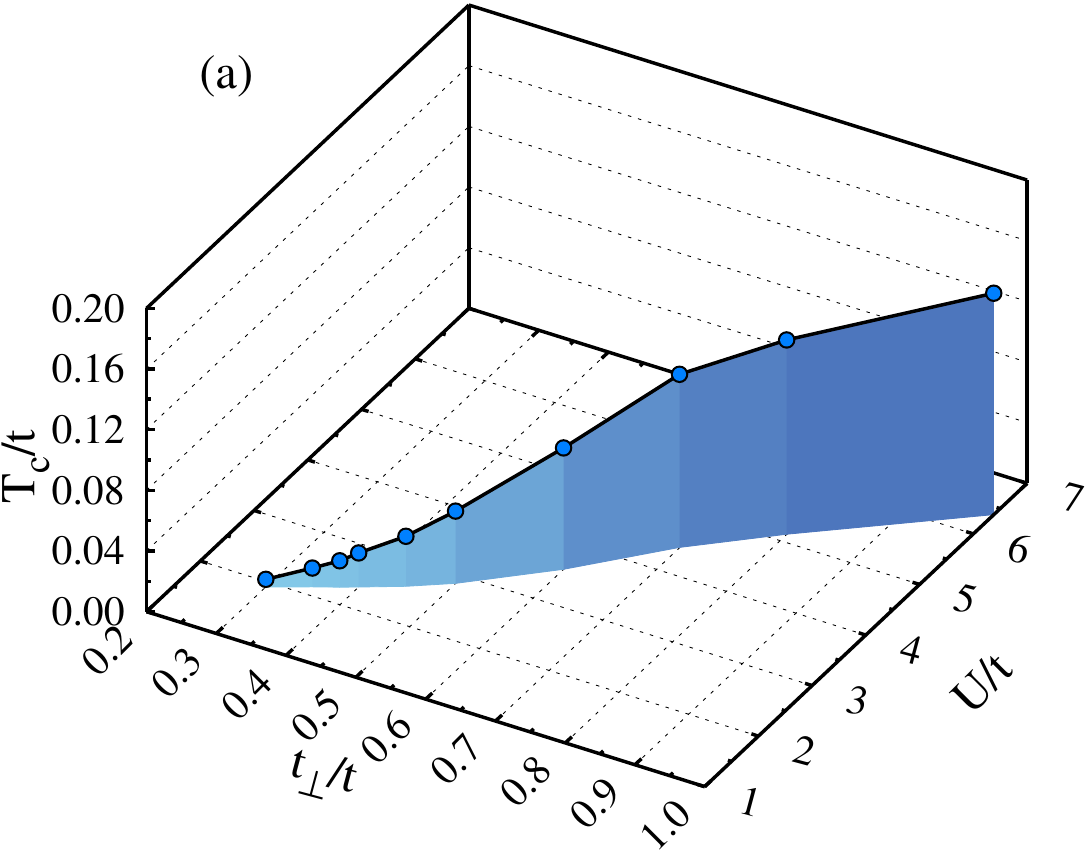}
\includegraphics[width=0.45\textwidth]{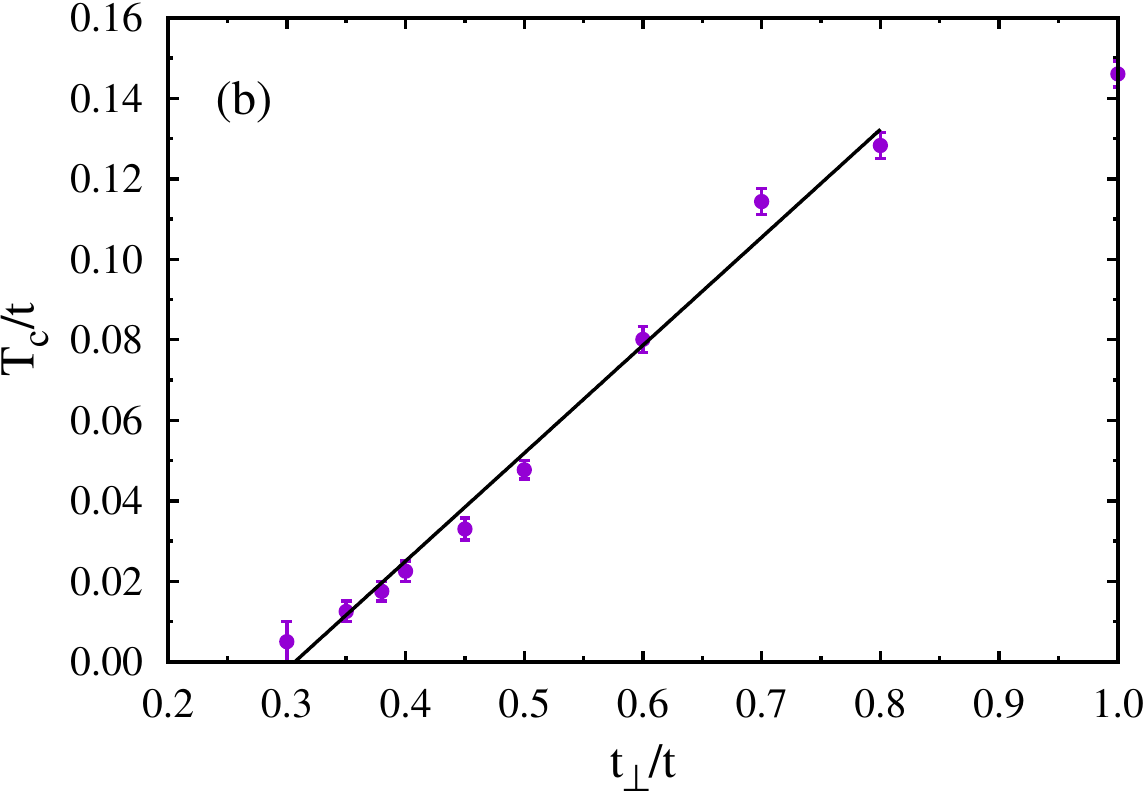}
\end{center}
\caption{(a) CDMFT estimate of the critical temperature and interaction strength $(T_c,U_c)$ 
terminating the first-order MIT extracted from the vanishing of the double occupancy jump; 
for $t_{\perp}/t= 0.3$ we could not see resolve any signature of the discontinuous behavior 
down to our lowest temperature $T=t/100$. 
(b) Linear fit to the data points in the $0.3\leqslant t_{\perp}/t\leqslant 0.8$ 
range which gives $t_{\perp}^c/t=0.31\pm0.03$.  
}
\label{T_c_fig}
\end{figure}

\subsection{\label{T_c} Critical end point $T_c$}

In this section we provide a detail analysis of the critical end point $(U_c,T_c)$ terminating the first-order MIT as a function of 
$t_{\perp}$.  Identifying  $(U_c,T_c)$  for a given $t_{\perp}$ requires numerous simulations at a variety of temperatures and as 
a function of $U$. We reduced this numerical effort by identifying first a crude estimate of the critical end point from 
simulations on a rough grid of temperatures.  Next, we pinpointed $(U_c,T_c)$ to a better degree of accuracy 
by performing additional simulations on an appropriately refined grid restricted to the vicinity of the critical end point. 

Figure~\ref{mag0853} collects the resultant data for the AF order parameter $m$ and the double occupancy $D$
for three representative values of $t_{\perp}$ corresponding  to different parts of the phase diagram. 
They range from $t_{\perp}/t=0.8$ [Figs.~\ref{mag0853}(a) and \ref{mag0853}(b)] through $t_{\perp}/t=0.5$ 
[Figs.~\ref{mag0853}(c) and \ref{mag0853}(d)] to $t_{\perp}/t=0.3$ [Figs.~\ref{mag0853}(e) and \ref{mag0853}(f)].  
As apparent, for both $t_{\perp}/t=0.8$ and $t_{\perp}/t=0.5$ one finds a temperature range where the continuous behavior 
of $m(U)$ and $D(U)$ changes into a discontinuous jump. 
In contrast, for $t_{\perp}/t=0.3$ we observe the persistence of a smooth behavior down to our lowest temperature $T=t/100$.

This motivated us to repeat the above analysis for other values of $t_{\perp}$ with the aim of elucidating 
the evolution of $T_c$ as a function of the interchain hopping. For each $t_{\perp}$ we define $T_c$ as a midpoint 
between the temperature at which $D(U)$ develops a singular behavior and the adjacent lower $T$ where $D(U)$ displays a clear jump. 
A striking outcome of this elaborate analysis is a nearly linear dependence of  $T_c$ versus $t_{\perp}$, see Fig.~\ref{T_c_fig}. 
A linear fit to the data points in the $0.3\leqslant t_{\perp}/t\leqslant 0.8$ range yields an estimate of a critical hopping 
$t_{\perp}^c/t=0.31\pm0.03$ at which the MIT ceases to be first order, see Fig.~\ref{T_c_fig}(b).

This result can be rationalized by invoking a basic concept behind the DMFT algorithm~\cite{Georges96}.
It describes the formation of renormalized quasiparticles as a self-consistent Kondo screening 
of local moments by the electronic bath. In the case of a single-site DMFT,  this screening 
involves only local spin-flip processes while in CDMFT also nonlocal spin-flips contribute.  
In addition, Kondo screening  competes with the AF superexchange interaction between local moments. 
Given a substantial reduction of the double occupancy down to $D=0.1348(4)$ in the PM metal on the verge of the MIT in Fig.~\ref{PD_cdmft}(b), 
indicative of well formed local moments $\langle S_z^2\rangle=1-2D$, this competition is expected to be particularly strong 
in the isotropic 2D limit.  In this limit we understand the Mott transition as a consequence of the jump to 
small concentration of 
doubly occupied sites (doublons) and empty sites (holons) triggered by the strong attraction of the doublon and holon scaled by $U$, 
which also drives the first-order transition due to the holon-doublon binding as described by the DMFT  approximation.  
Magnetism is just a consequence of the zero  doublon and holon state that is thermodynamically  unstable to magnetic ordering. 
In this sense  magnetism rides on the  MIT. This is very different  from the MIT in semimetals (i.e.  Hubbard model on a Honeycomb lattice)  
where  the charge  gap is a reflection of the magnetic ordering as in the case of Slater insulator~\cite{PhysRevX.3.031010,Raczkowski20}. 

Upon growing lattice anisotropy, the observed increase in $D$ on the metallic side of the MIT indicates that the system 
crosses over to a weak-coupling regime. This diminishes the impact of local moments on the nature of the MIT 
and reduces continuously the magnitude of a jump in $D$, and thus $T_c$,  down to zero.

\subsection{\label{Spectra} Electronic properties}

As discussed in Sec.~\ref{diagram}, lattice anisotropy controls the behavior of the doubly occupancy and hence the degree of localization 
in the metallic phase. 
This shall have a strong impact on electronic properties of the metal. Below we systematically  analyze the evolution of both Fermi surface and 
momentum-resolved single-particle spectra  $A(\pmb{k},\omega)$. It allows us to reveal dynamical effects arising from quantum fluctuations 
and to identify the underlying physics in different parts of the phase diagram. 

\subsubsection{\label{iso} Quasi-2D region: local moment formation}

\begin{figure}[t!]
\begin{center}
\includegraphics[width=0.2\textwidth]{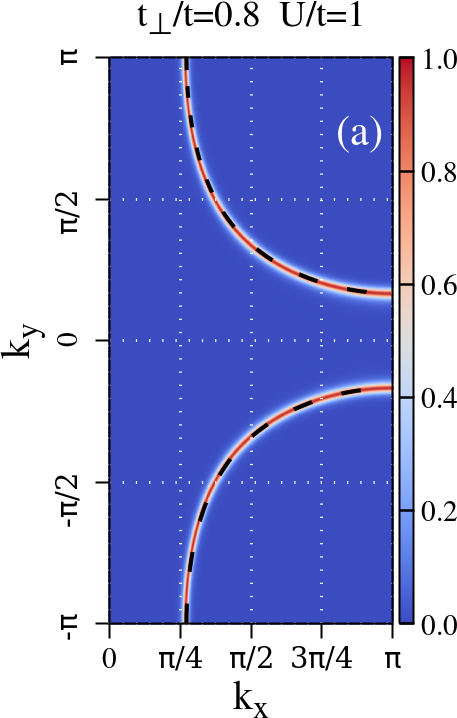}
\includegraphics[width=0.2\textwidth]{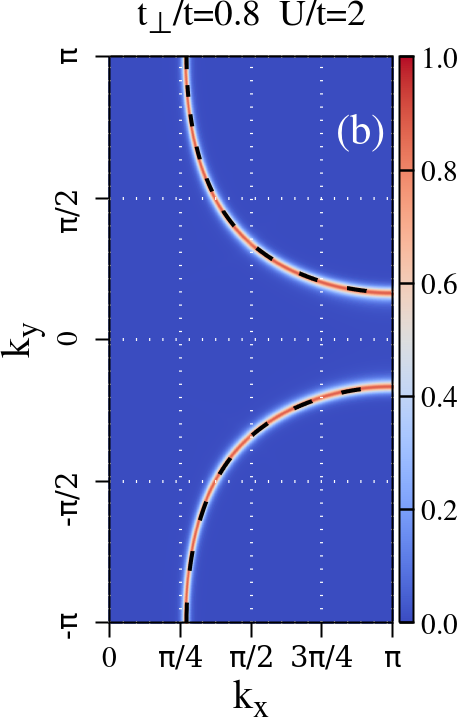} \\
\includegraphics[width=0.2\textwidth]{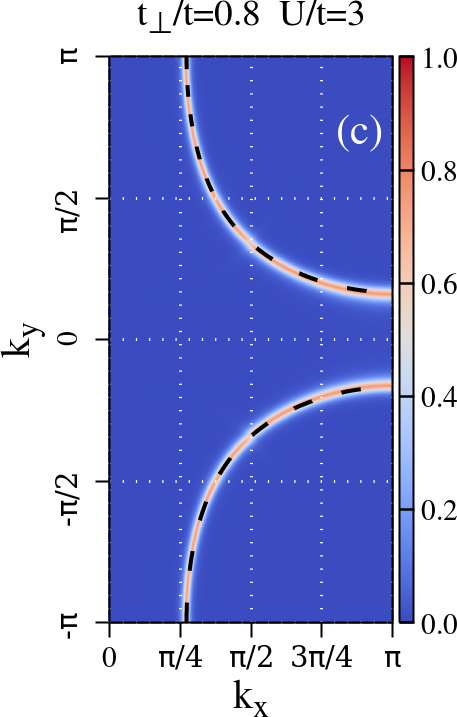}
\includegraphics[width=0.2\textwidth]{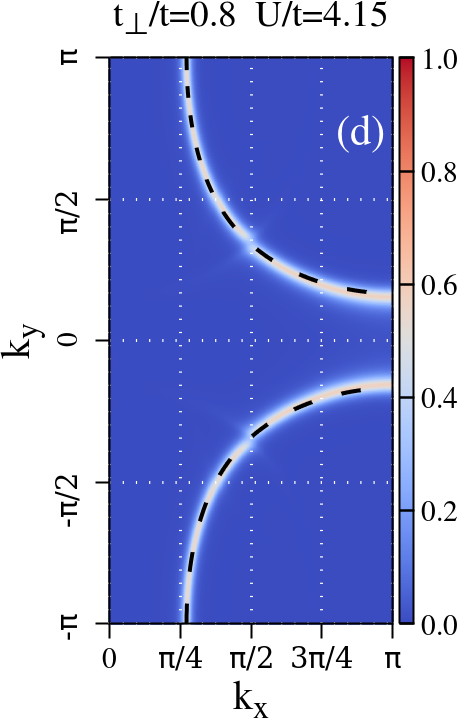}
\end{center}
\caption
{Evolution of the Fermi surface cuts across two quadrants of the Brillouin zone with increasing $U$ at $t_{\perp}/t=0.8$ 
in the PM phase at $T=t/40$. 
The dashed black line shows the non-interacting Fermi surface. 
The second Fermi surface segment in the Brillouin zone corresponds to  a mirror image about the $y$-axis.}
\label{FS_08}
\end{figure}

\begin{figure*}[t!]
\begin{center}
\includegraphics[width=0.4\textwidth]{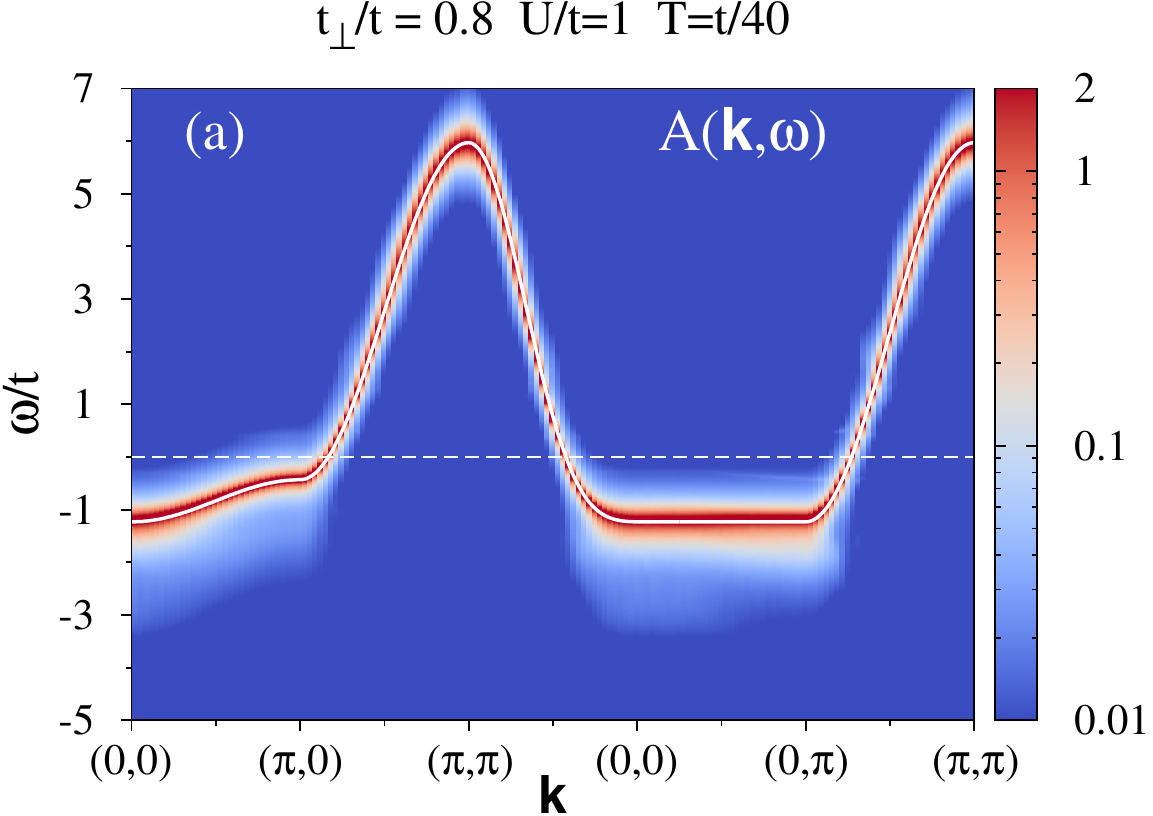}
\includegraphics[width=0.4\textwidth]{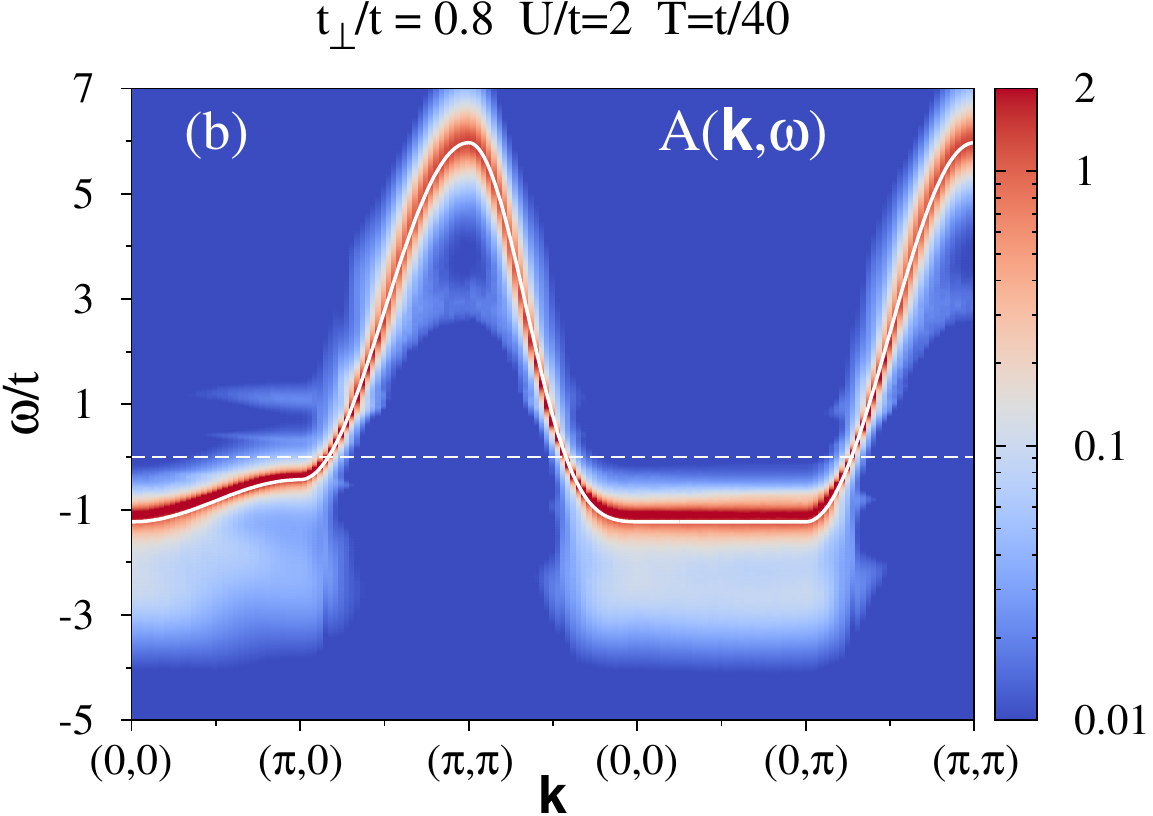}\\
\includegraphics[width=0.4\textwidth]{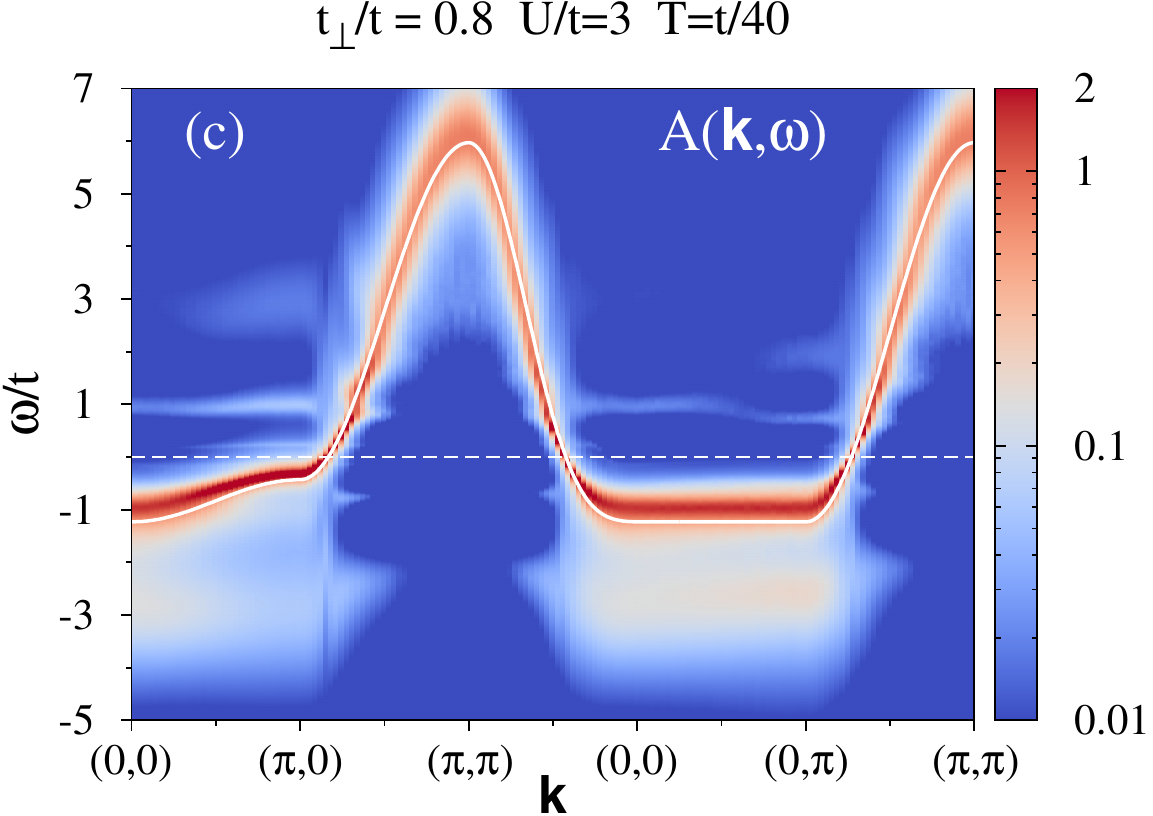} 
\includegraphics[width=0.4\textwidth]{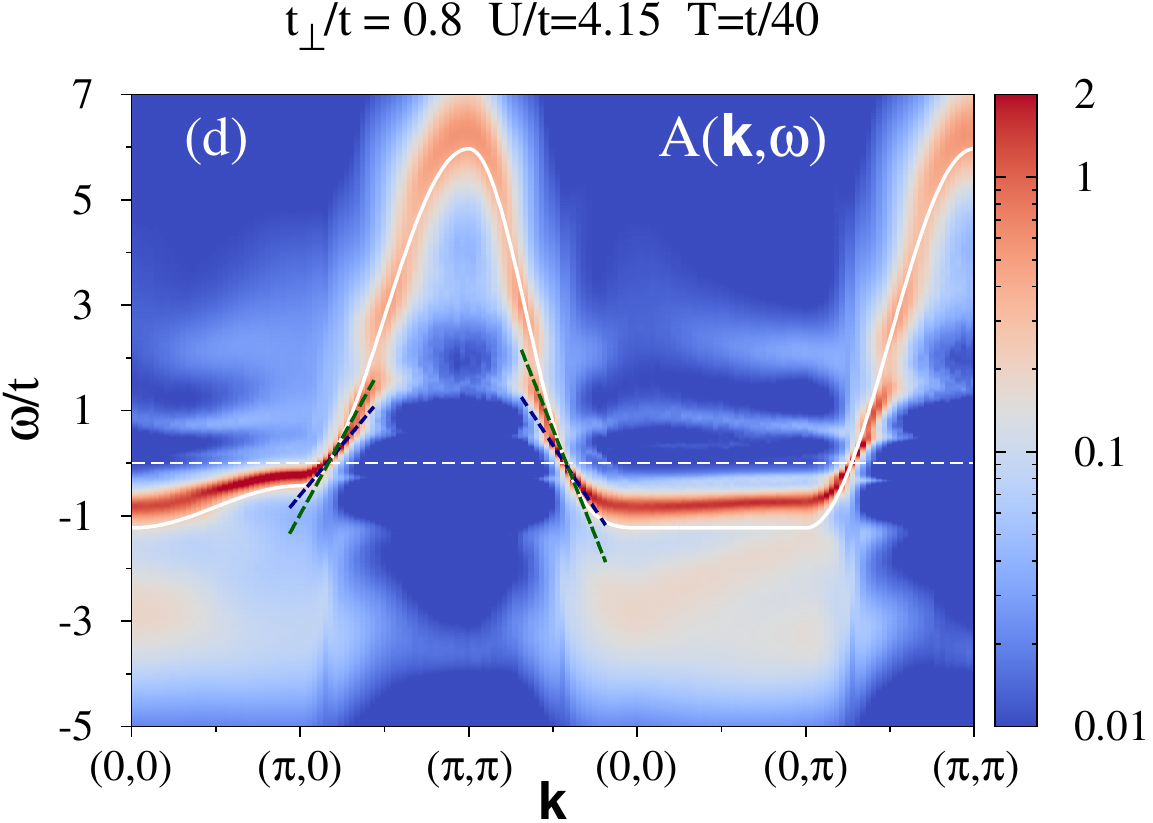}
\end{center}
\caption
{Single-particle spectrum $A(\pmb{k},\omega)$ with increasing $U$ 
at  $t_{\perp}/t=0.8$  in the PM phase at $T=t/40$. The solid white line shows the free dispersion. 
Dashed blue (green) line with weaker (steeper) slope in (d) denotes the Fermi velocity for the interacting (free) case, respectively.}
\label{Akw_08}
\end{figure*}

We begin with a weakly anisotropic case with  $t_{\perp}/t=0.8$. Figure~\ref{FS_08} displays the evolution of the Fermi surface segment across 
two quadrants of the Brillouin zone with increasing $U$ in the PM phase at constant temperature $T=t/40$. 
The dynamical contribution to the self-energy in CDMFT  yields a finite lifetime of quasiparticles. As a result, one finds a relatively sharp 
Fermi surface only for the smallest value $U/t=1$, see Fig.~\ref{FS_08}(a), while dynamical effects become  already discernible at $U/t=2$ 
as Fermi surface blurring, see Fig.~\ref{FS_08}(b). As shown in Figs.~\ref{FS_08}(c) and \ref{FS_08}(d), further increase of the interaction 
strength $U$ induces substantial transfer of spectral weight from the Fermi level to finite-frequency parts of the single-particle spectrum 
indicative of a correlated metal.

A few additional comments are in order: 

(i) It is known that in small cluster the effects of periodic boundary conditions are particularly strong which typically results in 
some artificial features in the single-particle spectra. For example, in Fig.~\ref{FS_08}(d) there is a noticeable depletion of 
weight at $k_x=\pi/2$. However, as we show later, its position is $t_{\perp}$ independent and pinned to $k_x=\pi/2$, and thus we consider 
it merely as a spurious feature of the $2\times 2$ cluster.

(ii) Since the $2\times 2$ CDMFT captures  short-range  AF spin fluctuations,  
the imaginary part of the self-energy can acquire a strong momentum dependence. If that is the case, the disappearance 
of the Fermi surface starts near the so-called hot spots, i.e.,  regions with an enhanced quasiparticle scattering rate. 
This gives rise to a pseudogap in the single-particle spectrum that precedes the Mott-Hubbard 
MIT~\cite{Parcollet04,Civelli05,Kyung06,Macridin06,Stanescu06,Zhang07,Tahara08,Park08,Werner09}. 
One could argue that the absence of momentum selective opening of the charge gap up to $U/t=4.15$, see Fig.~\ref{FS_08}(d), 
where the system is on the verge of the transition to the AF insulator, is simply because the critical interaction strength 
is smaller when AF spin order is allowed. However, we believe that it is the consequence of a large next-nearest-neighbor 
hopping $|t'|=t_{\perp}/2$ used in the present studies which brings about a strong frustration of the AF spin correlations. 
Because of the first-order nature of the transition, the transition takes place before an appreciable momentum dependence manifested 
by the momentum differentiation gets started when $U$ is increased. 
This conclusion is supported by previous studies of the 2D half-filled Hubbard model within the eight-site dynamical cluster approximation 
which reported the suppression of momentum-space differentiation as the magnitude of $t'$ increases~\cite{Gull09}.

(iii) Dynamical correlations can also renormalize the Fermi surface topology via the real part of the self-energy~\cite{Tocchio12}.  
Nevertheless, the inspection of Fig.~\ref{FS_08} indicates that increasing $U$ does not modify noticeably the Fermi surface shape 
such that it continues to follow the non-interacting one. 
Furthermore, anisotropic hopping amplitudes $t_{\perp}\neq t$ breaks the fourfold rotational $(C_4)$ symmetry.  This can lead to a 
surprisingly large directional anisotropy in the spectral intensity reflecting a large dynamically generated anisotropy in the self-energy 
close to the Mott-Hubbard MIT~\cite{Okamoto10}. 
We do not observe here such anomalies possibly due to a combined effect of a relatively small critical interaction sufficient 
to trigger the transition from a PM metal to the AF insulator and a strong magnetic frustration.

In order to gain further insight into the onset of a correlated metal, we plot in Fig.~\ref{Akw_08} the corresponding  
momentum-resolved single-particle spectra $A(\pmb{k},\omega)$. On the one hand, the spectral function in a weak-coupling regime $U/t=1$ 
follows essentially the non-interacting dispersion shown as the solid white line in Fig.~\ref{Akw_08}(a).
On the other hand,  qualitative changes in the spectrum  produced by dynamical correlations 
are already found at $U/t=2$: apart from the overall broadening, weak renormalization effects near the $\pmb{k}=(0,0)$ momentum 
become visible as a reduced bandwidth of the coherent quasiparticle dispersion  with respect to the non-interacting one, 
see Fig.~\ref{Akw_08}(b).
As shown in Figs.~\ref{Akw_08}(c) and \ref{Akw_08}(d), further increase in $U$ leads to the transfer of the zero-frequency spectral weight 
into higher-frequency regions as already anticipated in Fig.~\ref{FS_08}.
As a result, one observes the formation of the incoherent lower and upper Hubbard bands: 
the former appears predominantly in the region of Brillouin zone around the $\pmb{k}=(0,0)$ momentum while the latter emerges 
as $\pmb{k}$ moves towards the $(\pi,\pi)$ point. At the same time, the flattening of the quasiparticle dispersion  near the Fermi level  
signals an increased effective mass of the quasiparticles and thus growing localization tendency. 
To quantify this effect we have plotted in Fig.~\ref{Akw_08}(d) the  Fermi velocity  for the non-interacting and interacting cases. 
We see a  reduction of approximately 30\%    around $(\pi,0)$ and 40\%  around $(\pi/2,\pi/2)$ momenta.

\subsubsection{\label{melt} Quasi-2D region: thermal melting of local moments}

\begin{figure}[t!]
\begin{center}
\includegraphics[width=0.4\textwidth]{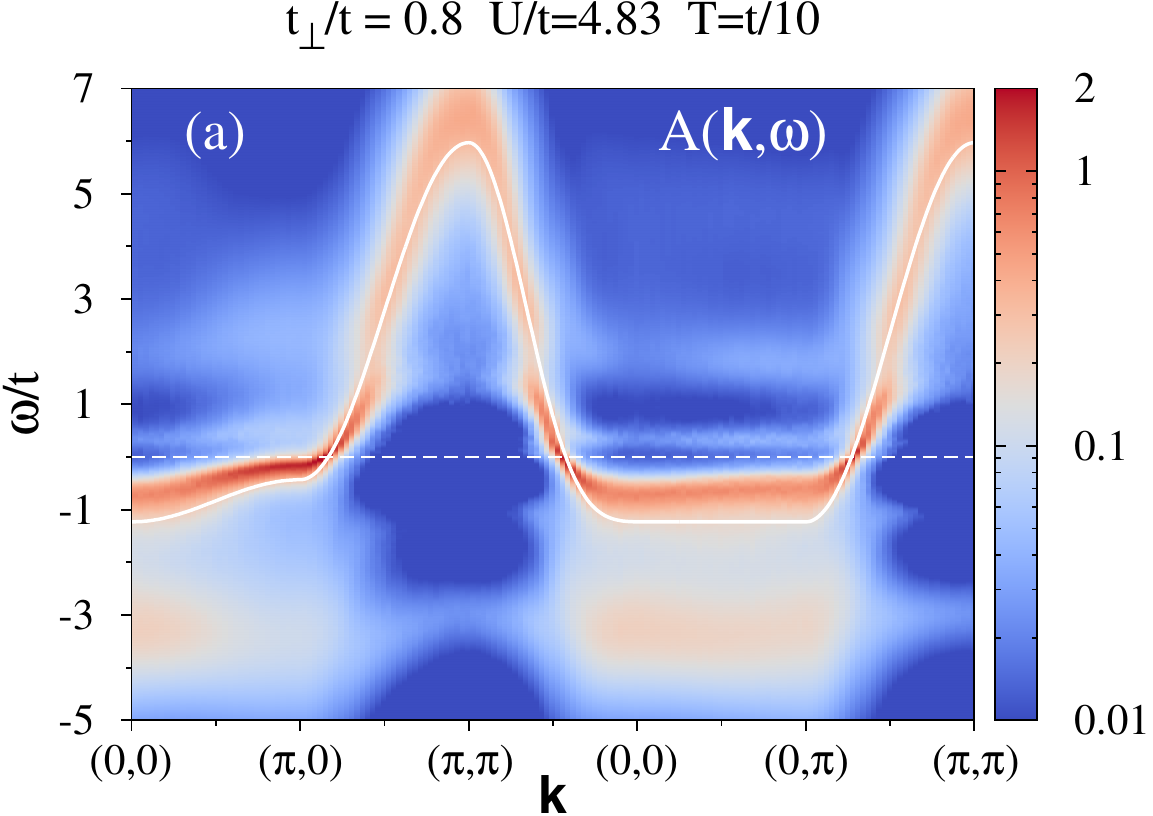}
\includegraphics[width=0.4\textwidth]{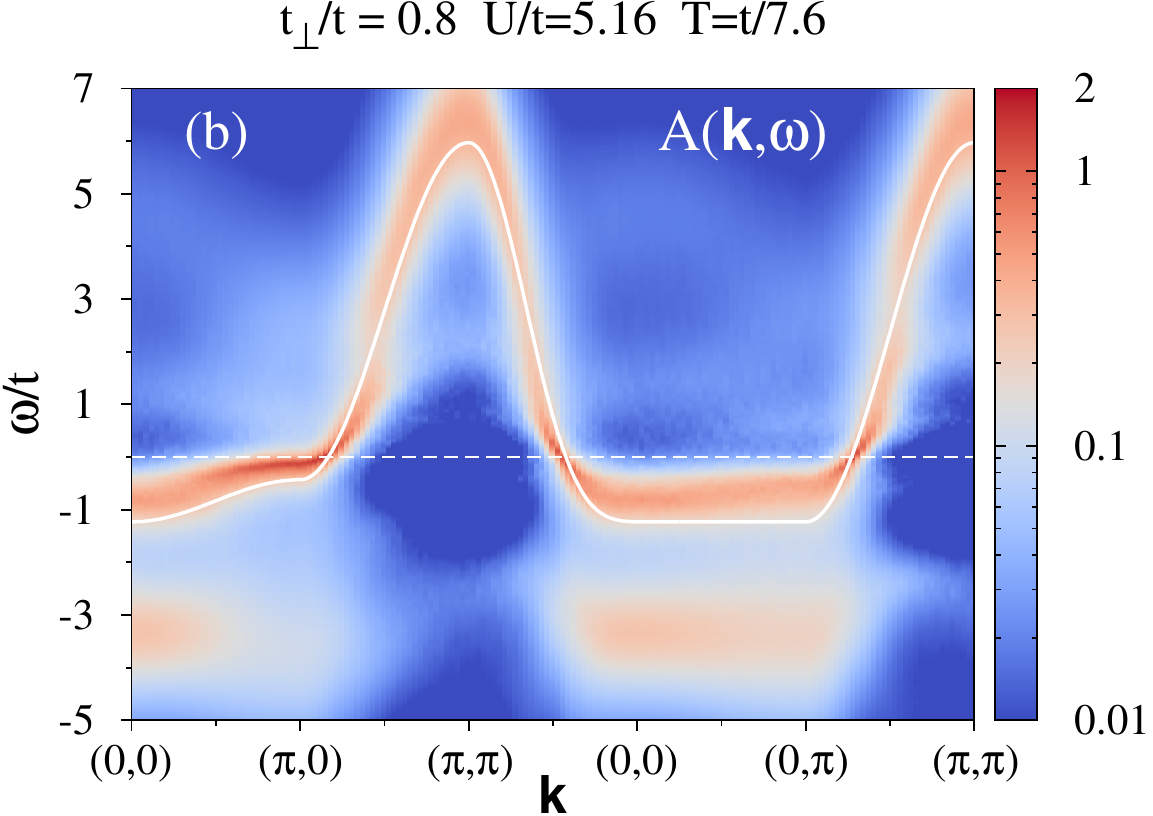}
\includegraphics[width=0.4\textwidth]{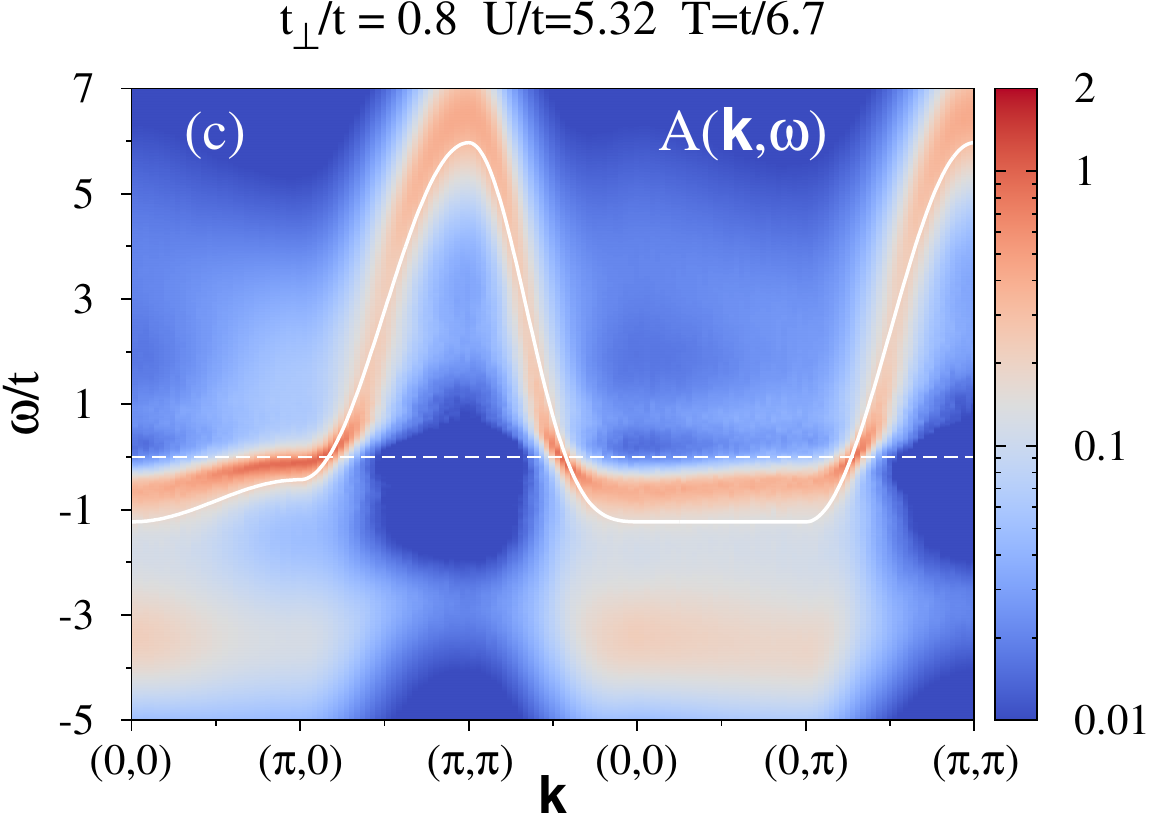}
\end{center}
\caption
{Evolution of the spectral function $A(\pmb{k},\omega)$ with increasing temperature $T$ at $t_{\perp}/t=0.8$ 
and at the critical interaction $U\lesssim U_c$ in the PM phase.}
\label{Akw_08T}
\end{figure}

A jump in the double occupancy across the MIT implies its first-order nature. 
It is thus natural to expect that the physical mechanism that underlies the opening of the charge gap, i.e., the local moment formation, 
is robust in temperature. 

We confirm this in Fig.~\ref{Akw_08T} by examining the temperature evolution of $A(\pmb{k},\omega)$ for a fixed $t_{\perp}/t=0.8$ 
and  at the corresponding critical interaction $U\lesssim U_c$ in the PM phase.
Indeed,  one finds the persistence of the coherent quasiparticle band near the Fermi level coexisting with incoherent high-frequency 
parts of the spectrum up to $T=t/10$, see Fig.~\ref{Akw_08T}(a). This is in accordance with a sizeable double-occupancy jump 
at this temperature illustrated in Fig.~\ref{mag0853}(b). 
Indeed, in the strongly correlated metal with $U/t\gg 1$, the high-energy scale $\sim U$ determines spectral properties already in the
high-temperature regime $T\simeq  U$. It leads to the formation of incoherent high-energy features (which are precursive of the lower 
and upper Hubbard bands in the Mott insulator) in addition to the quasiparticle peak at the Fermi level. 
In the DMFT picture of the first-order metal-insulator transition below $T_c$, spectral weight is transferred from the zero-frequency 
quasiparticle peak to (already preformed) high-frequency features and the transition is signaled by a jump in double occupancy.

At even higher $T=t/7.6$ [Fig.~\ref{Akw_08T}(b)], i.e., the highest $T$ 
at which one observes a jump in the double occupancy, a remnant quasiparticle band is still resolved in a narrow frequency window 
near the Fermi level. 
Finally, at our highest $T=t/6.7$ [Fig.~\ref{Akw_08T}(c)] the low-frequency quasiparticle band is washed out but the high-frequency 
features continue to be visible. At this temperature $T>T_c$, we could only detect a smooth transition from the PM to AF phase, 
see Figs.~\ref{mag0853}(a) and \ref{mag0853}(b). 
That is in accord with our line of arguing that the incoherent high-frequency features should stay intact across the
critical end point $T_c$ since they start to form already at higher temperature $\simeq U$.

\subsubsection{\label{moderate} Moderate anisotropy: Fermi-surface topology change}

\begin{figure}[t!]
\begin{center}
\includegraphics[width=0.4\textwidth]{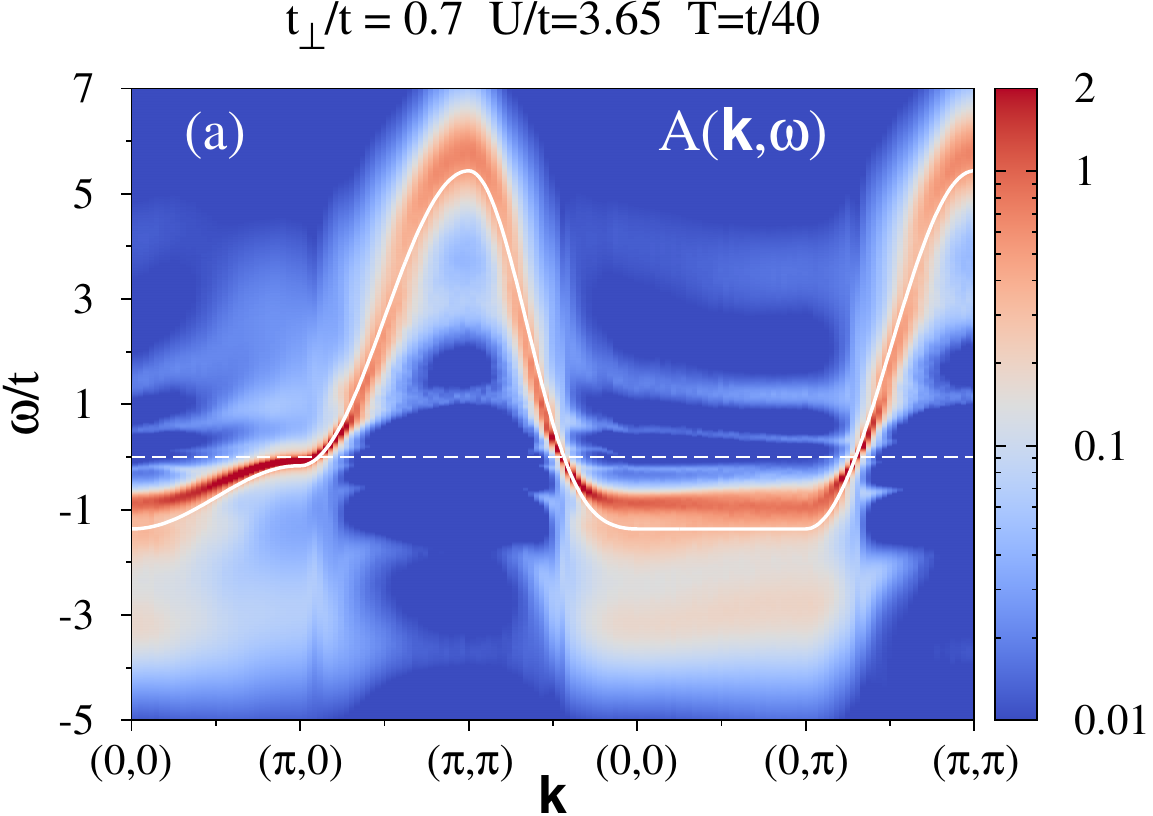}
\includegraphics[width=0.4\textwidth]{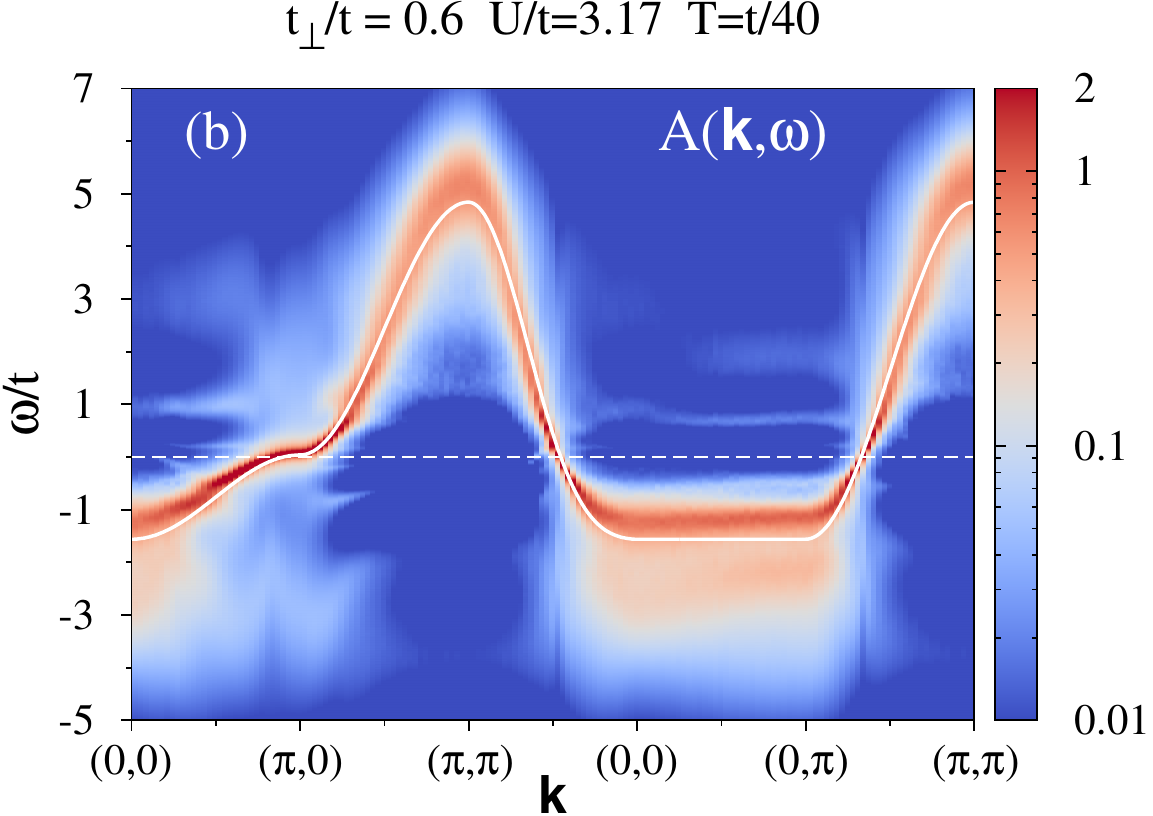}
\end{center}
\caption
{Spectral function $A(\pmb{k},\omega)$ at the critical interaction $U\lesssim U_c$  in the PM phase at $T=t/40$ 
in the moderately anisotropic region: (a) $t_{\perp}/t=0.7$ and (b) $t_{\perp}/t=0.6$.}
\label{Akw_076}
\end{figure}

The presence of saddle points of energy dispersion yields a van Hove singularity in the single-particle density of states. 
Typically the access to study the behavior of a system near a van Hove singularity is achieved by the fine tuning of the electron 
density such that the Fermi surface approaches the singularity. 
Here we provide evidence that the lattice anisotropy is yet another control parameter that allows one to drive the van Hove singularity 
to cross the Fermi level. 

\begin{figure}[t!]
\begin{center}
\includegraphics[width=0.2\textwidth]{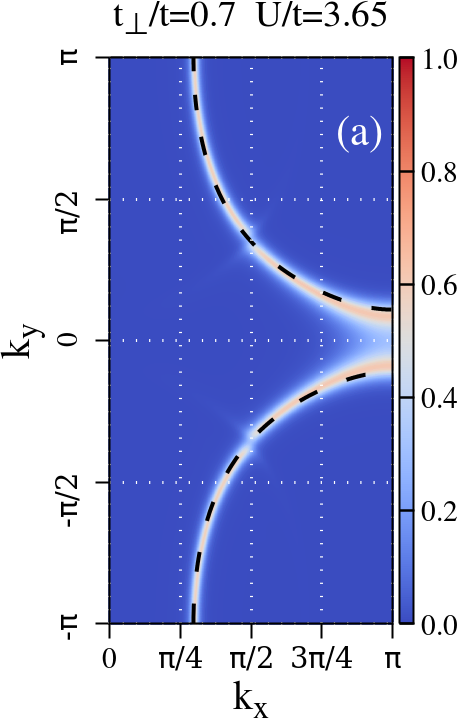} 
\includegraphics[width=0.2\textwidth]{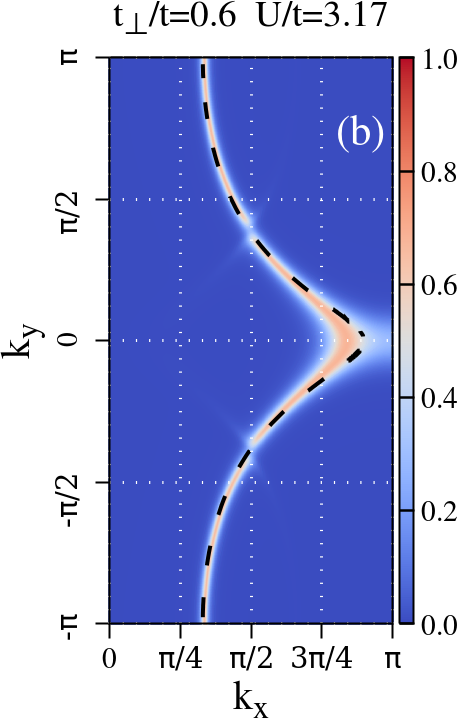}
\end{center}
\caption
{Topological (Lifshitz) transition of the Fermi surface:  at $t_{\perp}/t=0.7$ (a) the Fermi surface is closed  around the Brillouin zone corner 
at $\pmb{k}=(\pi,\pi)$ while at $t_{\perp}/t=0.6$ (b) one finds an open quasi-1D Fermi surface. 
In both panels, the value of $U$ corresponds to the critical interaction $U\lesssim U_c$ in the PM phase at $T=t/40$ while
the dashed black line shows the non-interacting Fermi surface.
} 
\label{FS_076}
\end{figure}

We illustrate it in Fig.~\ref{Akw_076} which shows $A(\pmb{k},\omega)$ at the critical interaction $U\lesssim U_c$ 
in the PM phase at $T=t/40$: a saddle-point region at ${\pmb k}=(\pi,0)$ located below the Fermi level at $t_{\perp}/t=0.7$, 
see Fig.~\ref{Akw_076}(a), crosses the Fermi level upon increasing the lattice anisotropy such that at $t_{\perp}/t=0.6$ it is found 
above the Fermi energy, see Fig.~\ref{Akw_076}(b). The flat dispersion crossing the Fermi level contributes low-energy states 
and gives rise to the enhanced density of states $N(\omega=0)$  found in Fig.~\ref{PD_cdmft}(c) in the range  $0.6\le t_{\perp}/t\le 0.7$.
One can also notice in Fig.~\ref{Akw_076}(b) that the incoherent spectral weight at high negative frequency occupies a narrow
energy range and its maximum moves towards the non-interacting dispersion.
Moreover, at the same $t_{\perp}$ where the saddle-point region at ${\pmb k}=(\pi,0)$  crosses the Fermi level,  the system  
undergoes a Lifshitz transition  whereby the Fermi surface topology changes from a closed to an open  one, see Fig.~\ref{FS_076}.   
Thus we confirm a one-to-one correspondence between this type of Lifshitz transition  and the van Hove singularity 
crossing the Fermi level~\cite{Chen12}.

\subsubsection{\label{aniso} Quasi-1D region: itinerant antiferromagnetism}

\begin{figure}[t!]
\begin{center}
\includegraphics[width=0.4\textwidth]{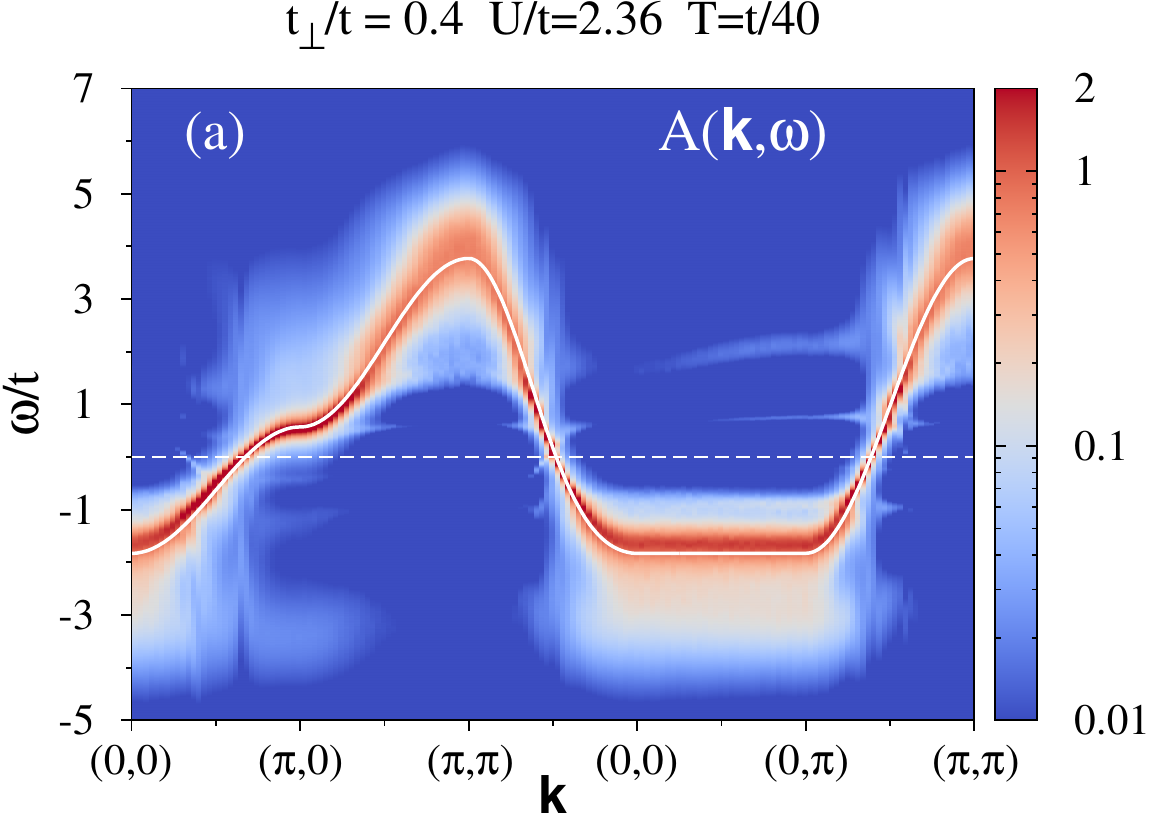}
\includegraphics[width=0.4\textwidth]{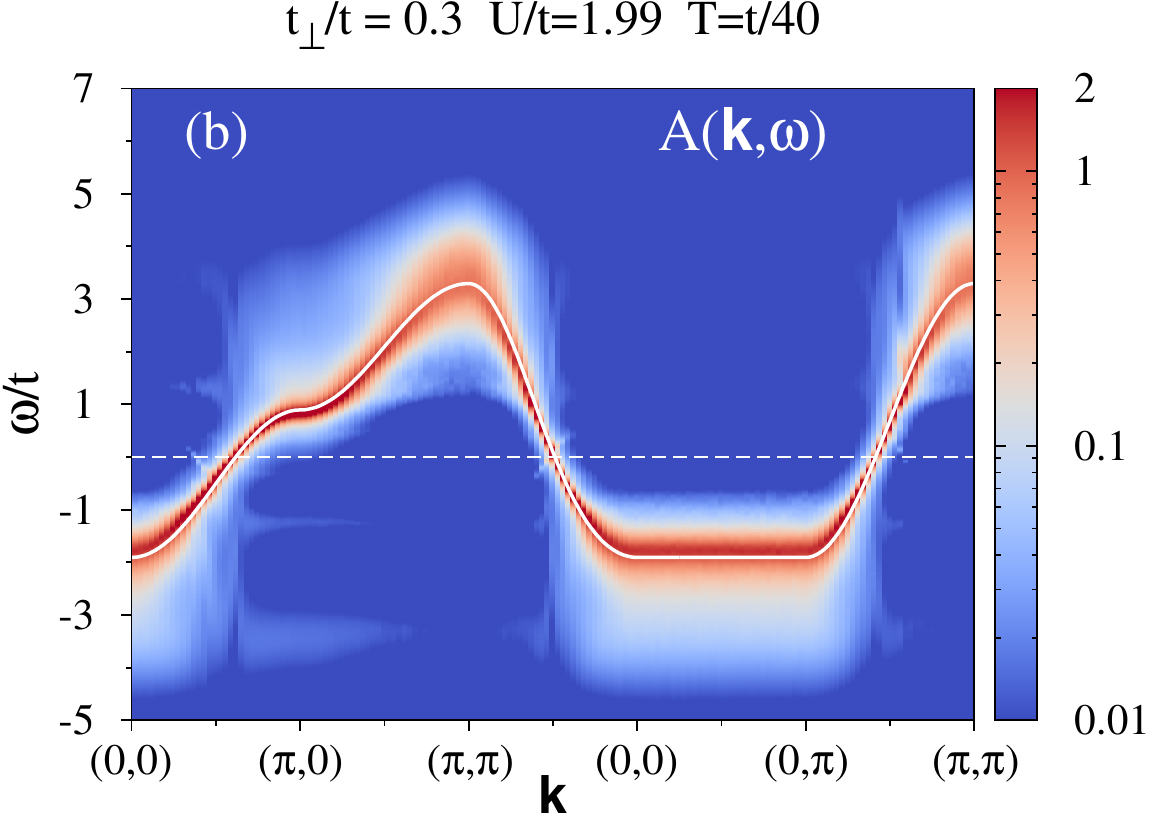}
\includegraphics[width=0.4\textwidth]{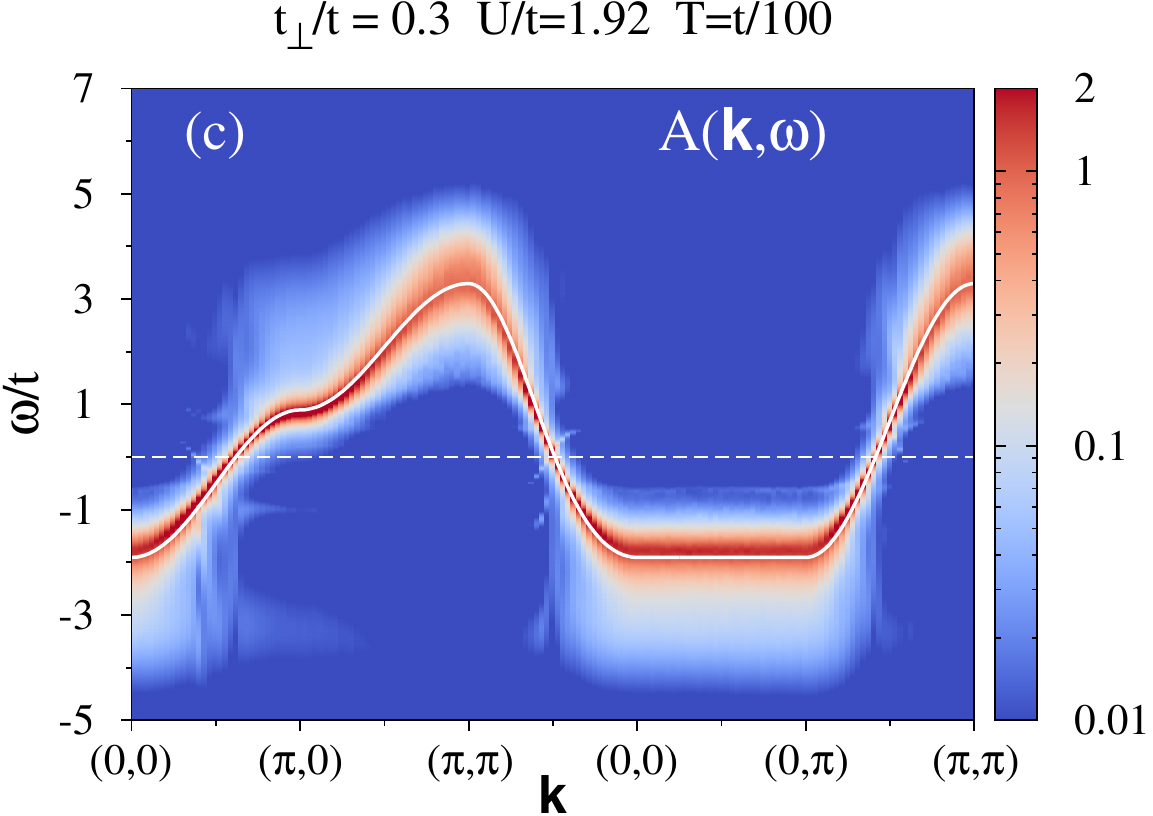}
\end{center}
\caption
{Spectral function $A(\pmb{k},\omega)$ at the critical interaction $U\lesssim U_c$  in the PM phase at $T=t/40$ 
in the strongly anisotropic region with: (a)  $t_{\perp}/t=0.4$ and (b) $t_{\perp}/t=0.3$.  
For comparison, panel (c) shows $A(\pmb{k},\omega)$ for $t_{\perp}/t=0.3$ at our lowest $T=t/100$.}
\label{Akw_043}
\end{figure}

The vanishing critical end point $T_c$ identified in Sec.~\ref{T_c} suggests that local temporal fluctuations are not anymore the primary 
mechanism driving the localization on a strongly anisotropic lattice. 
Namely, a well established hallmark of DMFT is that by taking into account local temporal fluctuations, it reproduces a three-peak 
spectrum (lower Hubbard band, quasiparticle peak, upper Hubbard band) of a strongly correlated metal.
At the critical $U$, the metal-insulator transition is signaled then by the disappearance of the quasiparticle peak and the transition 
is found in DMFT to be first order below the critical end point $T_c$.
The vanishing $T_c$ is the quasi-1D region indicates that there must be a different mechanism of localization at play, otherwise 
the transition would continue to be of first order. Indeed, as we discuss below,  
we find there a continuous splitting of the quasiparticle band in the single-particle spectrum 
(which closely resembles, as we show in Fig.~\ref{Akw_043}, that of the free electrons) due to doubling of the unit cell in the AF phase.

First of all, let us recall a relatively large double occupancy on the PM side of the phase diagram in Fig.~\ref{PD_cdmft}(b),   
ranging from $D=0.1795(2)$ at $t_{\perp}/t=0.4$ to $D=0.1978(1)$ at $t_{\perp}/t=0.2$ on the verge of the magnetic transition.  
This has a direct impact on the resultant ${\pmb{k}}$-resolved spectral function. 
As apparent in Fig.~\ref{Akw_043}(a), already at $t_{\perp}/t=0.4$, the position of the intensity maximum for 
a given momentum $\pmb {k}$ matches rather well the non-interacting dispersion. Meanwhile, it is  only at high frequency  where 
$A(\pmb{k},\omega)$ displays some broadening which can be considered as remnants of the two Hubbard bands. 
The same observation  holds true for $A(\pmb{k},\omega)$ at $t_{\perp}/t=0.3$. 
To exclude the possibility  that a close resemblance between the CDMFT and non-interacting  spectra  at $T=t/40$ 
results merely from dominant thermal effects, 
we display in Fig.~\ref{Akw_043}(c) $A(\pmb{k},\omega)$ at $t_{\perp}/t=0.3$ at our lowest temperature  $T=t/100$.  
The absence of any emerging correlation-driven effects in the spectrum, which continues to follow the non-interacting dispersion relation, 
confirms the irrelevance of a local moment physics and provides  further support in favor of quantum critical behavior  
below a critical anisotropy $t_{\perp}^c/t=0.31\pm0.03$ as established in Fig.~\ref{T_c_fig}(b).

In the above  statement we make use of the fact that the double occupancy $D$ is
a measure of the correlation strength: in the non-interacting limit $U = 0$,
$D$ takes its uncorrelated value 1/4. As $U$ grows, $D$ decreases until it is fully
suppressed which corresponds to the spin-1/2 Heisenberg limit.
The single-particle spectral function in the PM phase depends then on whether the numerical approach 
captures the aforementioned reduction of double occupancy.
At the static mean-field level, local moments cannot be generated without
breaking the SU(2) spin symmetry as double occupancy continues to keep the
uncorrelated value 1/4. Consequently, spectral function in the PM phase is identical to the non-interacting one.
In contrast, the DMFT approximation accounts for local electronic correlations
which systematically reduce the double occupancy even before the magnetic
transition takes place. That is reflected in the redistribution of spectral
weight and leads to the onset of high-energy features coexisting with the low-energy quasiparticle band 
as we illustrate it in Fig.~\ref{Akw_08} for $t_{\perp}/t=0.8$.
However, for small $t_{\perp}/t=0.3$, we observe that the reduction of double occupancy is much weaker. 
It  matches the absence of the extra high-energy features in the corresponding spectral function in Figs.~\ref{Akw_043}(b) and \ref{Akw_043}(c).
As such it is essentially accounted for by a non-interacting dispersion.
Hence, the PM phase in this range of $t_{\perp}/t$ does not feature well formed local moments and can be described by weak-coupling approaches.

\begin{figure}[t!]
\begin{center}
\includegraphics[width=0.2\textwidth]{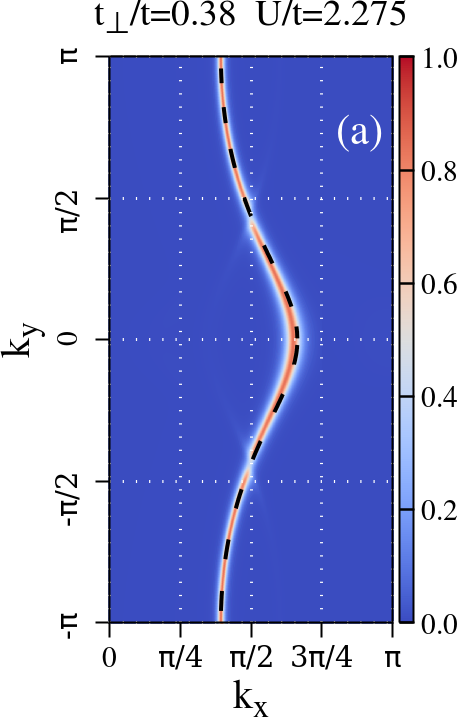}
\includegraphics[width=0.2\textwidth]{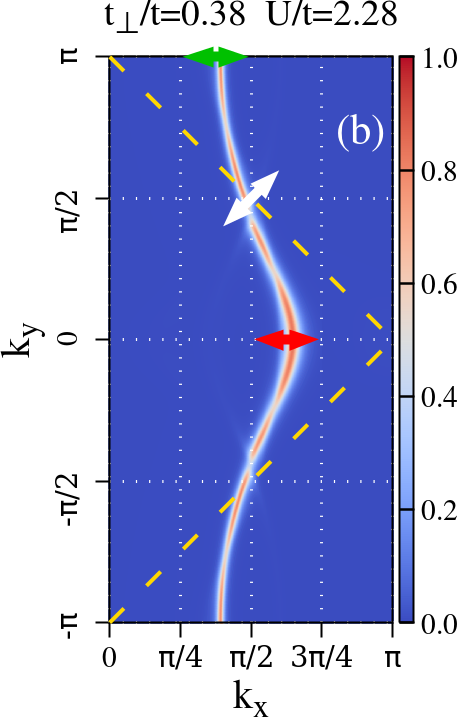}\\
\includegraphics[width=0.4\textwidth]{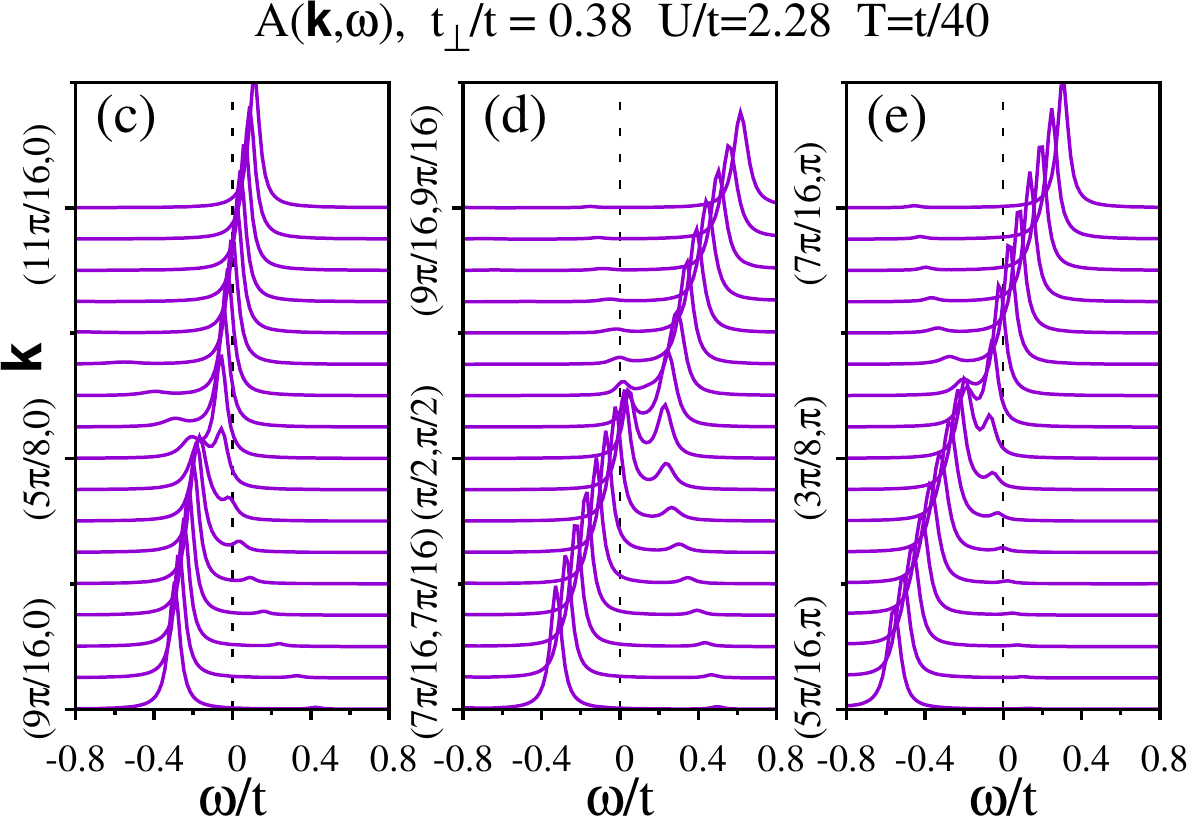}\\
\includegraphics[width=0.4\textwidth]{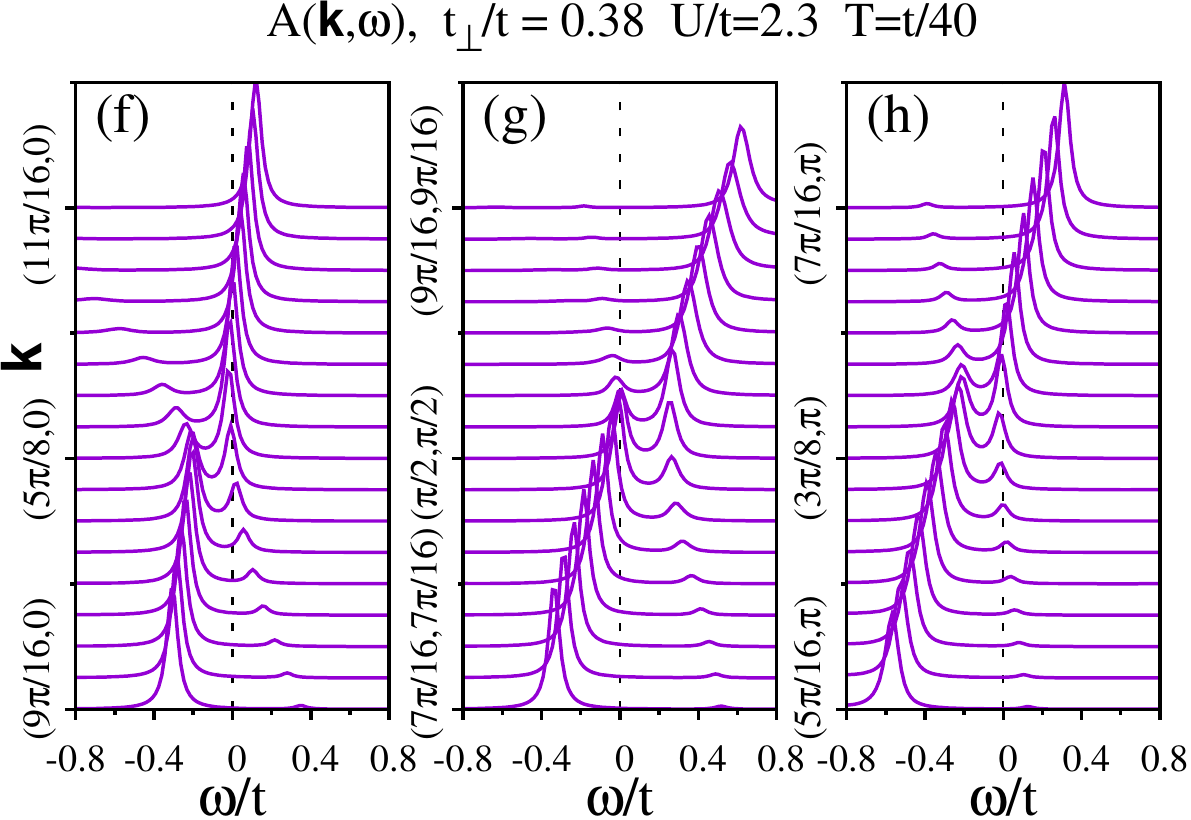}\\
\end{center}
\caption
{Single-particle properties in the proximity of magnetic transition for $t_{\perp}/t=0.38$ at $T=t/40$.
Fermi surface cuts on the: (a) PM and (b)  AF side of the transition. In (b) the measured value of the staggered magnetization $m=0.0749(6)$. 
As a consequence of relatively weak magnetic order, the whole Fermi surface arc continues to exist on the AF side with a partial 
suppression of the spectral weight at the hot spots in the AF phase.
(c)-(e) Low-energy part of the spectral function $A(\pmb{k},\omega)$   along
$(0,0)\to(\pi,0)$, $(0,0)\to(\pi,\pi)$,  and $(0,\pi)\to(\pi,\pi)$ paths in the AF metal 
at $U/t=2.28$ and (f)-(h) at $U/t=2.3$ with the magnetization  $m=0.1419(4)$.
Red, white, and green arrows in (b) indicate the actual momentum range shown in panels (c) and (f), (d) and (g),  (e) and (h), respectively, 
while the dashed yellow line shows the AF Brillouin zone.
}
\label{sp_038}
\end{figure}

\begin{figure}[t!]
\begin{center}
\includegraphics[width=0.2\textwidth]{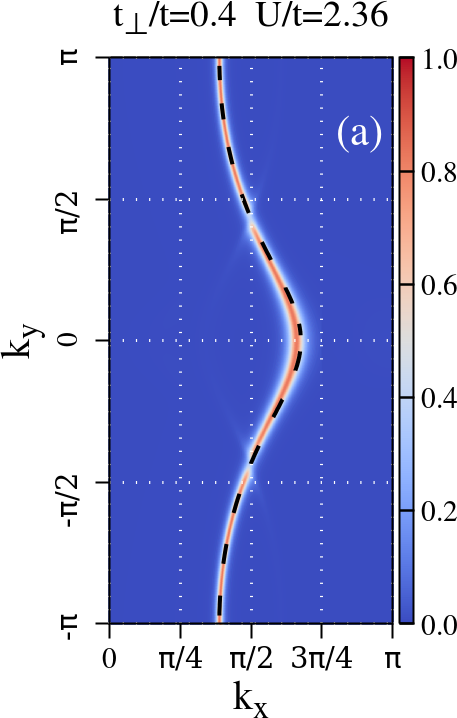}
\includegraphics[width=0.2\textwidth]{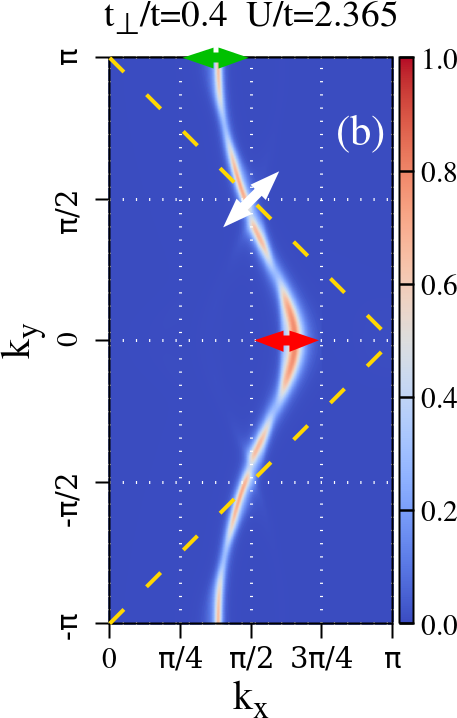}\\
\includegraphics[width=0.4\textwidth]{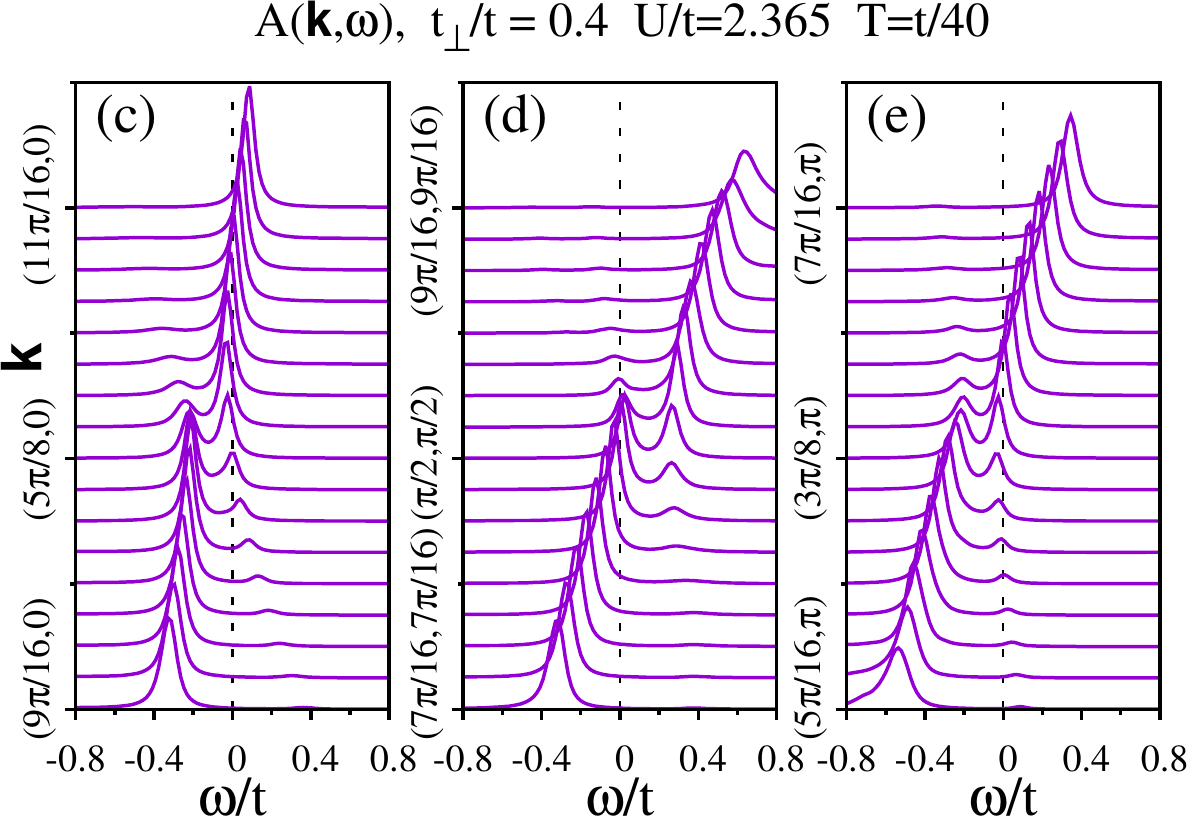}\\
\includegraphics[width=0.4\textwidth]{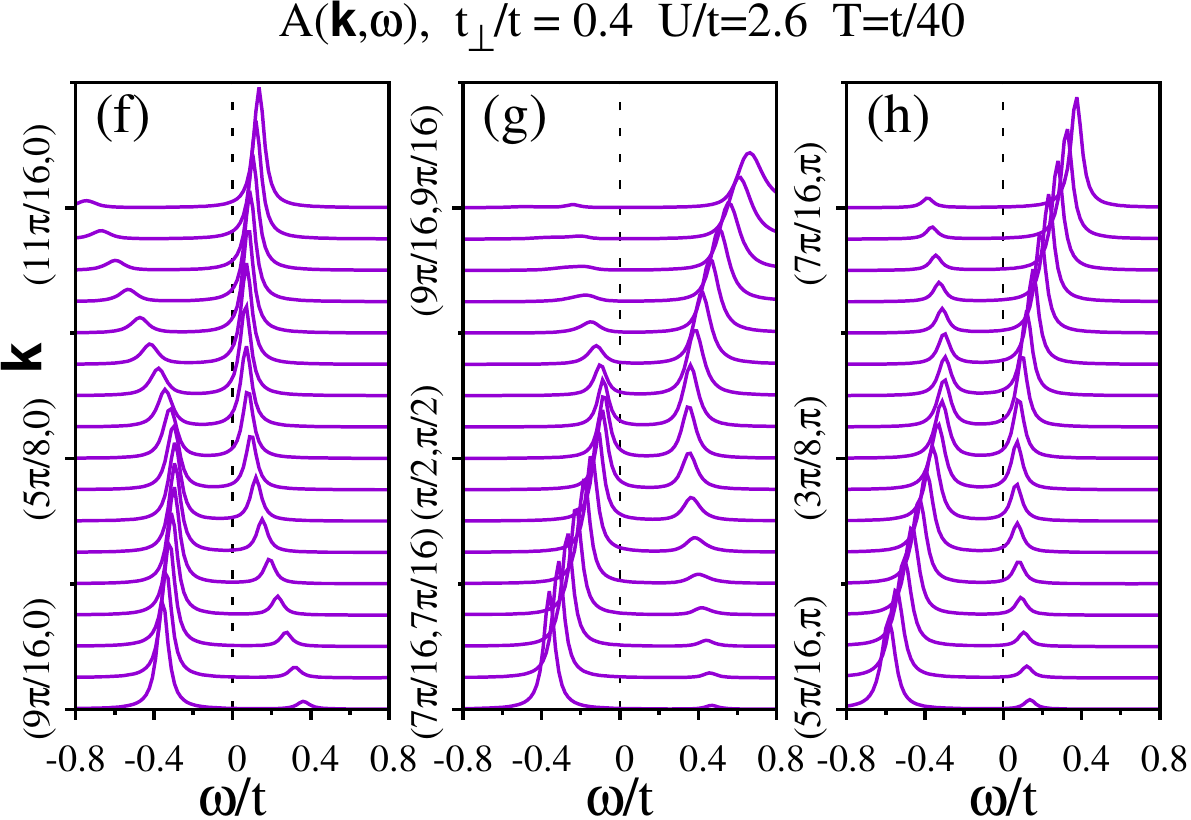}
\end{center}
\caption
{Single-particle properties in the proximity of magnetic transition for $t_{\perp}/t=0.4$ at $T=t/40$: 
Fermi surface cuts  on the: (a) PM and (b) AF side of the transition. In (b) the measured value of the staggered magnetization $m=0.1275(8)$. 
Due to strong AF order,  the Fermi surface arc breaks into disconnected  pockets. 
(c)-(e) Low-energy part of $A(\pmb{k},\omega)$  along the Brillouin zone paths as in Fig.~\ref{sp_038} in the AF metal at $U/t=2.365$ 
and (f)-(h) in the AF insulator at $U/t=2.6$.
}
\label{sp_04}
\end{figure}

Another signature of a reduced quasiparticle scattering off local moments at $t_{\perp}/t=0.4$ is substantially restored coherence 
of low-energy quasiparticle excitations.  Indeed, the resultant distinct Fermi surface found for both $t_{\perp}/t=0.38$ 
[Fig.~\ref{sp_038}(a)] and $t_{\perp}/t=0.4$ [Fig.~\ref{sp_04}(a)] contrasts sharply with that found 
at $t_{\perp}/t=0.8$ where a clear loss of spectral weight is apparent, cf. Fig.~\ref{FS_08}(d).
However, a precise physical mechanism of the magnetic phase transition for $t_{\perp}/t=0.38$ differs 
at $T=t/40$ from that for $t_{\perp}/t=0.4$.  We discuss now both cases separately.  

In accordance with a continuous nature of the transition at $t_{\perp}/t=0.38$,   
one observes in the AF phase  a smooth disappearance of the Fermi surface near "hot" regions, see Fig.~\ref{sp_038}(b). 
A more detailed inspection of the low-frequency part of $A(\pmb{k},\omega)$ on the AF side of transition shows 
the backfolding of the quasiparticle dispersion due to the broken translation symmetry and the resultant depletion of spectral 
weight just below (above) the Fermi level in Figs.~\ref{sp_038}(c) and \ref{sp_038}(e) [Fig.~\ref{sp_038}(d)], respectively.
This depletion should be considered as  a precursive feature of electron  and hole pockets that open up at larger $U$.
Indeed,  one finds  that the backfolded  quasiparticle band crosses the Fermi level at two momenta --- the spectral weight at the second crossing
is much weaker than the original quasiparticle dispersion and causes a faint "ghost" side of the pockets.
In fact, the observed gradual reconstruction of the low-energy quasiparticle dispersion can be reproduced by  a functional form
\begin{equation}
	E^{\pm}_{\ve{k}} = \frac{ \epsilon_{\ve{k} } + \epsilon_{\ve{k}+\ve{Q} }} {2}  
		\pm\sqrt{  \left(\frac{ \epsilon_{\ve{k}} - \epsilon_{\ve{k}+\ve{Q} } } {2} \right)^2   + \Delta^2},
\end{equation}
where $\Delta=Um/2$, consistent with that of the mean-field band structure in the spin-density-wave state. 
Hence, we identify the origin of the itinerant antiferromagnetism as Slater-like with quasiparticle scattering off essentially static 
staggered moment whose growing magnitude controls the size of the pockets as $U$ grows, see Figs.~\ref{sp_038}(f)-\ref{sp_038}(h).

\begin{figure}[t!]
\begin{center}
\includegraphics[width=0.4\textwidth]{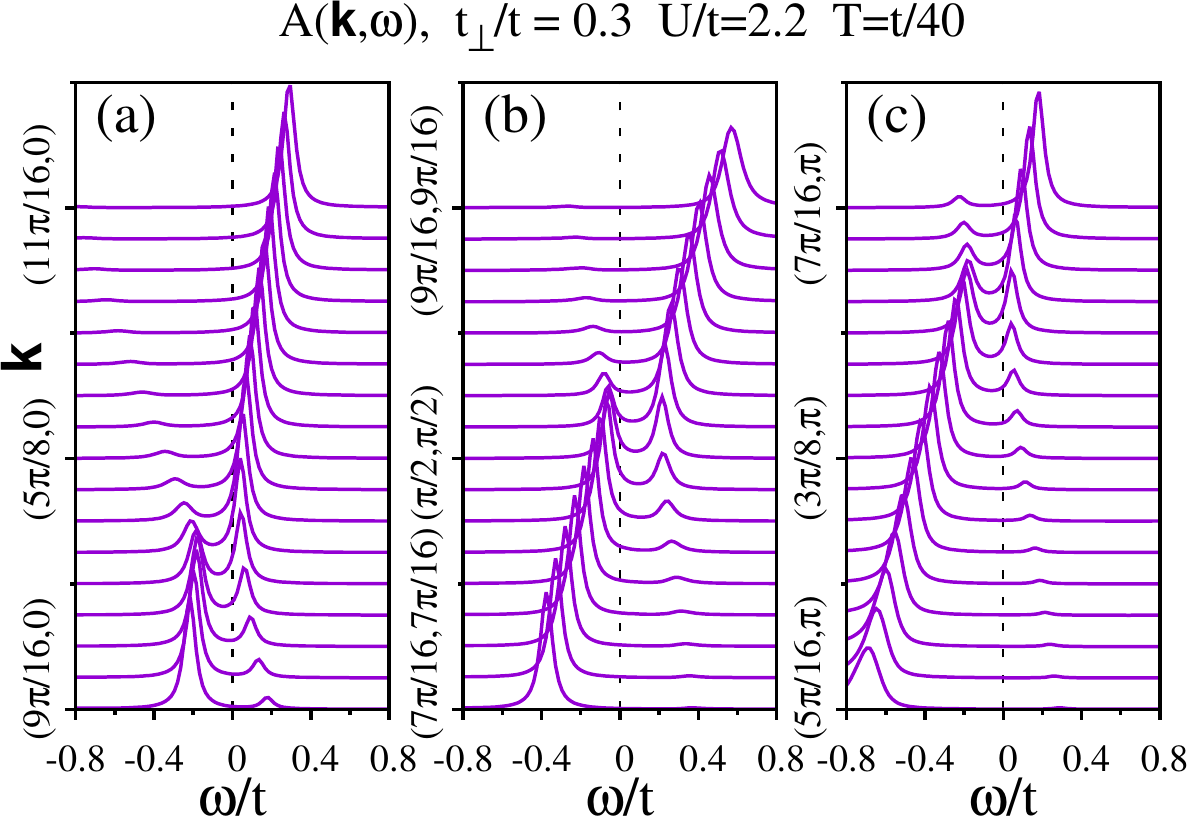}\\
\includegraphics[width=0.4\textwidth]{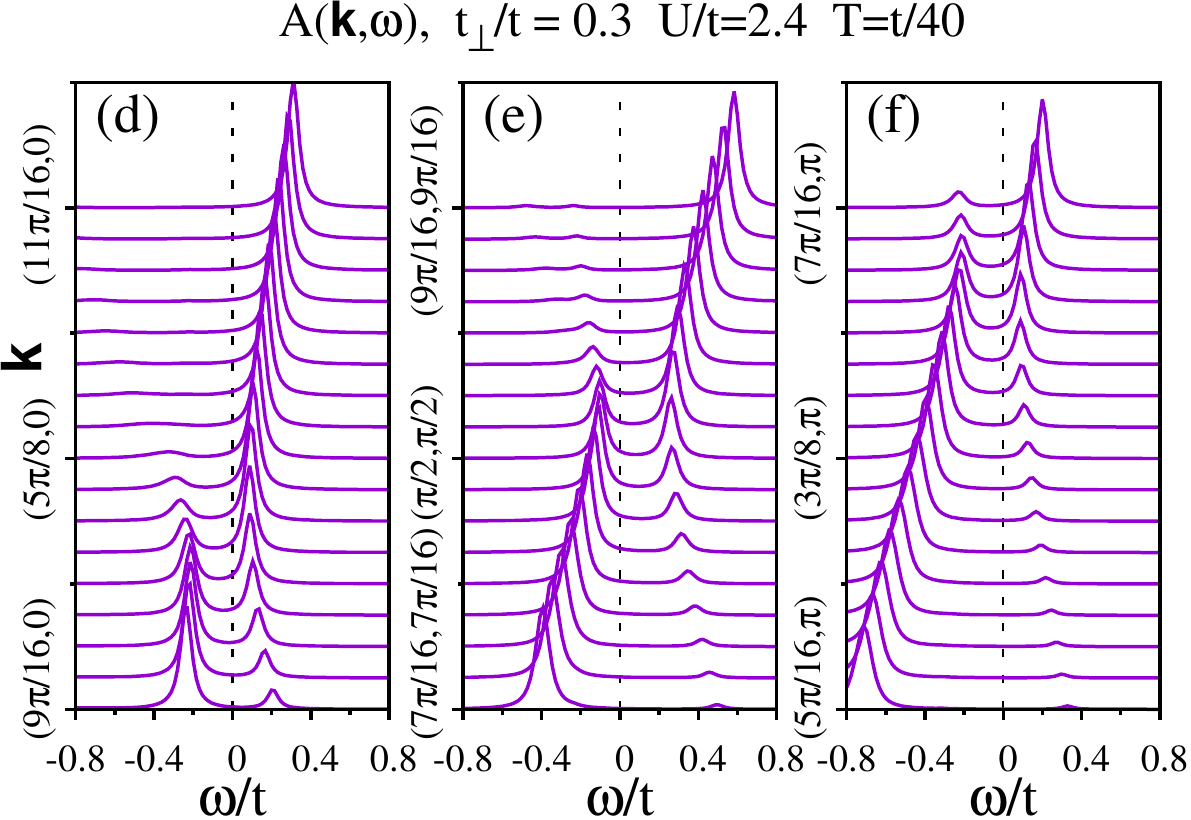}
\end{center}
\caption
{Low-energy part of the spectral function $A(\pmb{k},\omega)$ at $t_{\perp}/t=0.3$ and $T=t/40$ along
$(0,0)\to(\pi,0)$ (a) and (d), $(0,0)\to(\pi,\pi)$ (b) and (e),  and $(0,\pi)\to(\pi,\pi)$ (c) and (f) paths in the Brillouin zone 
at $U/t=2.2$ (top) and $U/t=2.4$ (bottom)  across the transition from an  AF metal to the AF insulator.
}
\label{MIT_t03}
\end{figure}

\begin{figure}[t!]
\begin{center}
\includegraphics[width=0.4\textwidth]{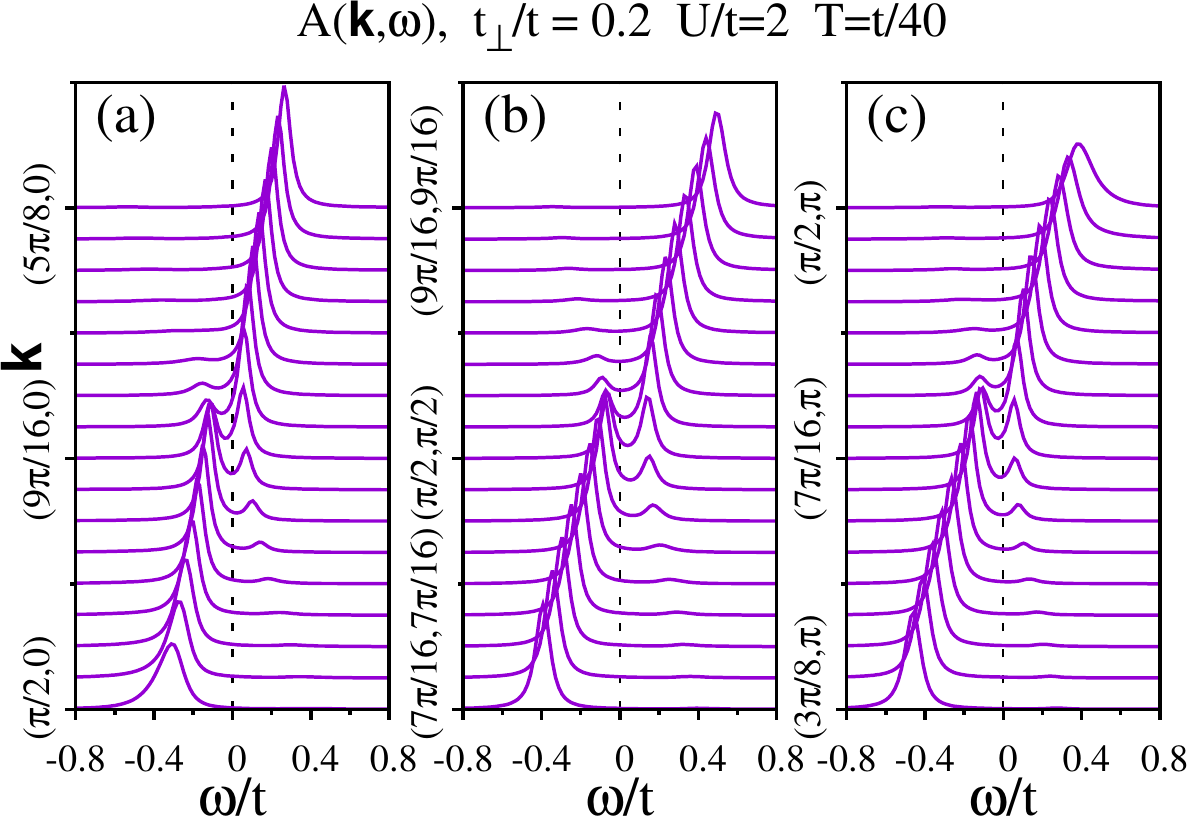}\\
\includegraphics[width=0.4\textwidth]{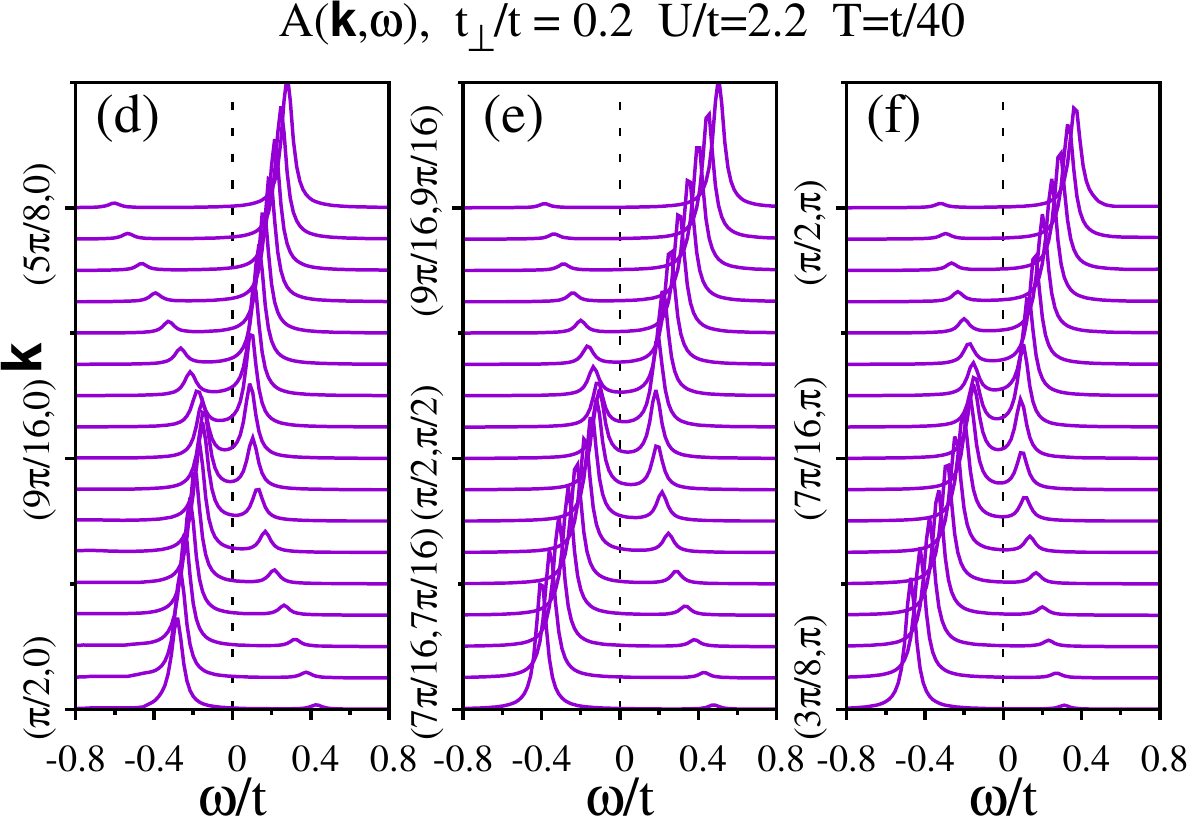}
\end{center}
\caption
{Same as in Fig.~\ref{MIT_t03} but for $t_{\perp}/t=0.2$ at $U/t=2$ (top) and $U/t=2.2$ (bottom).
}
\label{MIT_t02}
\end{figure}

In contrast,  the magnetic transition at $t_{\perp}/t=0.4$  is weakly first order and  significant AF order 
builds up right away at the critical interaction. This rapid develop is a consequence of a remnant local moment physics which 
enhances localization effects. This is reflected in disconnected Fermi surface segments on the AF side of the magnetic transition, 
see Fig.~\ref{sp_04}(b), accompanied by a definite gap between the lower and upper quasiparticle bands illustrated in 
Figs.~\ref{sp_04}(c)-\ref{sp_04}(e).  
Upon further increasing $U$, the bands shift away from the Fermi level, gradually reducing the  size of Fermi pockets 
in the intermediate AF metal phase which ultimately brings about the transition into the AF insulator, see Figs.~\ref{sp_04}(f)-\ref{sp_04}(h). 
Hence, we find aspects of both local and nonlocal correlation participating in the emergence of the insulating  phase.

We turn now to the analysis of the low-frequency part of $A(\pmb{k},\omega)$ as a function of $U$ for smaller hoppings  
 $t_{\perp}/t=0.3$ and $t_{\perp}/t=0.2$  with the goal of elucidating the location of the corresponding MITs. In each case, 
 a continuous loss of metalicity upon increasing $U$ is accompanied by the occurrence of more pronounced backfolded quasiparticle features 
 such that they are visible in a broader momentum range, see Figs.~\ref{MIT_t03} and \ref{MIT_t02}. 
Let us also point out that the identified redistribution of spectral weight restricted to  a narrow frequency region on the scale 
of the charge gap of the AF insulator is a generic feature of the magnetic instability in a weakly correlated 
metal driven predominately by the Fermi surface.  This contrasts with the strong coupling regime where 
the onset of AF order in the Mott insulator involves the spectral weight transfer within the two Hubbard bands 
such that it accumulates at their lower edges~\cite{Fleck04,Fratino17a}. Hence, the emergent magnetic ordering modifies 
the spectral properties  over a broad energy range that is much larger than the AF charge gap itself.

The resultant MIT phase boundary anticipated from Figs.~\ref{sp_04}, \ref{MIT_t03}, and \ref{MIT_t02} complements 
the CDMFT phase diagram shown in Fig.~\ref{PD_cdmft}.  
Together with the line indicating the onset of the staggered magnetization, they both delimit the domain of stability of 
itinerant antiferromagnetism. It matches qualitatively the region of phase space with strongly reduced but nevertheless 
finite $N(\omega=0)$, see Fig.~\ref{PD_cdmft}(c).

Let us conclude the discussion of the established AF metal phase by contrasting its Fermi surface topology 
with that obtained in the previous studies in Refs.~\cite{Raczkowski12,Lenz16} restricted to the normal phase of the 
model Eq.~(\ref{eq:Hubb}). In this case the destruction of the Fermi surface starts at momenta ${\pmb k} = (\pi/2,\pm\pi/2)$ 
where the interchain hopping matrix elements vanish and is driven by the remnant 1D umklapp scattering~\cite{Essler02,Penc11}. 
The resultant broken Fermi surface of the compensated metal displays elliptic electron and hole pockets around 
the ${\pmb k}=(\pi/2,0)$ and $(\pi/2,\pm\pi)$ points, see Fig.~5 of Ref.~\cite{Lenz16}.

\section{\label{discuss} Summary and conclusions}

Understanding the nature and factors controlling the degree of electron localization is crucial for exploring 
functional applications of quantum materials such as transition metal oxides. In this work we have made a contribution 
to this issue by  studying the interplay between electron correlation, frustration, and dimensionality effects 
in the anisotropic Hubbard model at half-filling. To this end, we have adapted CDMFT to handle long-range AF order.
An important outcome from our study is that the quasi-1D region of the magnetic phase diagram harbors an AF 
metal. Consequently, in the CDMFT scenario one finds a crossover from a local moment physics of a correlated 
isotropic 2D metal to the itinerant AF behavior in the strongly anisotropic case.

It is very interesting that, independently of a specific implementation of CDMFT, i.e., paramagnetic or broken spin symmetry, 
one observes a full suppression of the critical end point $T_c$ of the MIT upon approaching the quasi-1D region. 
In both cases, the emergent quantum criticality can be traced back to a growing relative importance  of spatial versus 
local fluctuations. Indeed,  strong quasiparticle scattering off local moments along the \emph{whole} Fermi surface explains its sudden 
disappearance  and the resultant first-order character of the MIT in the 2D case. On the contrary, damping of low-energy
quasiparticles in the quasi-1D region  begins near  "hot" regions of the Fermi surface. This leads to the formation of hole and electron 
Fermi surface pockets in a resultant compensated metal which ultimately undergoes a continuous MIT. 

Specifically, when the CDMFT loop is constrained to converge to the PM solution, the MIT is driven by remnant 1D umklapp scattering 
and corresponds to the vanishing of the Fermi pockets driven by their  continuous shift away from the Fermi level~\cite{Raczkowski12,Lenz16}.
Likewise, when AF spin order is allowed, imperfect nesting of the model band structure paves the way to the itinerant AF transition 
followed by a MIT whose continuous nature is again the consequence of a smooth disappearance of Fermi pockets. 
In the actual simulations we explicitly broke the translational symmetry of the lattice by allowing for 
a nonvanishing staggered magnetic moment. The latter is equivalent to divergence of the static spin susceptibility at the  
AF wavevector  $\pmb{Q}=(\pi,\pi)$ measured on a sufficiently large cluster size capturing correctly the correlation 
length scale of spin fluctuations. From this point of view it becomes clear that introducing the lattice anisotropy tips 
the balance between local temporal fluctuations responsible for the Mott-Hubbard physics and thus---the first-order MIT---and 
spatial AF spin fluctuations playing a key role in the established continuous transition from an AF metal to the AF insulator.

The established CDMFT phase diagram with PM, AF metal, and AF insulator phases bears similarity with that of the extended Hubbard model 
featuring at the mean-field level  PM, charge-ordered metal, and charge-order insulator phases~\cite{Imada06,Imada07}.
In the latter case one finds a tricritical point where all the three phases coexist. This very special point terminates also 
the continuous transition between the PM and charge-ordered metal phases. In our analysis we were unable to locate such 
a tricritical point. Instead as a function of the anisotropy $t_{\perp}/t$ we find three situations:  
(i) first-order transition from a PM metal to the AF insulator; 
(ii) a crossover region where the dynamical correlations trigger a weakly first-order  
AF transition but they are not strong enough to fully localize charge carriers, and 
(iii) continuous transition between the PM and AF metal phases.
Still, an attempt to induce the tricritical point could be made by fine tuning of the non-interacting band structure,
i.e, the ratio of $t'/t_{\perp}$.

Another possible scenario in the strongly anisotropic limit  would be the interaction-driven emergence of the genuine Mott insulator
without any symmetry breaking characterized as the quantum spin liquid and sandwiched by the PM metal and the AF insulator. 
Such an intervening Mott insulating quantum spin liquid phase is beyond the scope of the CDMFT approximation.
Since  the auxiliary-field quantum Monte Carlo algorithm is hindered in the presence of geometrical frustration by the negative sign 
problem~\cite{ALF2017}, a many-variable variational Monte Carlo method is an appealing option~\cite{VMC08}  to clarify this point of view.

Our second important result is evidence that the next-nearest-neighbor hopping $t'$ brings about an efficient mechanism to suppress 
the AF instability. In particular, we found that in the isotropic 2D situation with $t'=-t/2$,  the PM metal extends to a fairly 
large interaction $U/t=5.17$ below which previous CDMFT studies of the half-filled $t$-$t'$-Hubbard model reported a finite $d$-wave 
superconducting order parameter~\cite{Sentef11}. Since those studies did not consider long range AF order, it was possible that the 
latter prevails and leads to the insulating behavior at $T=0$ instead. Our findings corroborate the scenario that the frustration of AF 
spin interactions by finite $t'$ shifts the onset of antiferromagnetism to a critical interaction which is large enough to expose 
$d$-wave superconductivity~\cite{Raghu10,Eberlein14}.
Moreover, given our evidence for itinerant antiferromagnetism with Fermi surface pockets, it would be worth examining, e.g., 
using the spin fluctuation approach~\cite{Andersen16}, the leading superconducting pairing instabilities in the spatially anisotropic 
situation.

\begin{acknowledgments}
We would like to acknowledge enlightening discussions with B. Lenz, K. Takai, and Y. Yamaji.  
This work was supported by the German Research Foundation (DFG) through Grant No. RA 2990/1-1.
FFA acknowledges financial support from the DFG through the W\"urzburg-Dresden Cluster of Excellence on Complexity 
and Topology in Quantum Matter - ct.qmat (EXC 2147, project-id 39085490) as well as  through the SFB 1170 ToCoTronics.
MI  acknowledges the support by MEXT as "Program for Promoting Researches on the Supercomputer Fugaku"
(Basic Science for Emergence and Functionality in Quantum Matter and the HPCI project HP200132) and
Kakenhi (Grant No. 16H16345).
The authors gratefully acknowledge the Gauss Centre for Supercomputing e.V. (www.gauss-centre.eu) 
for funding this project by providing computing time through the John von Neumann Institute for Computing (NIC) 
on the GCS Supercomputer JUWELS~\cite{juwels}   at J\"ulich Supercomputing Centre (JSC).
\end{acknowledgments}

\appendix

\section{\label{app:sign} Negative sign problem of the CT-QMC cluster solver}

\begin{figure}[b!]
\begin{center}
\includegraphics[width=0.4\textwidth]{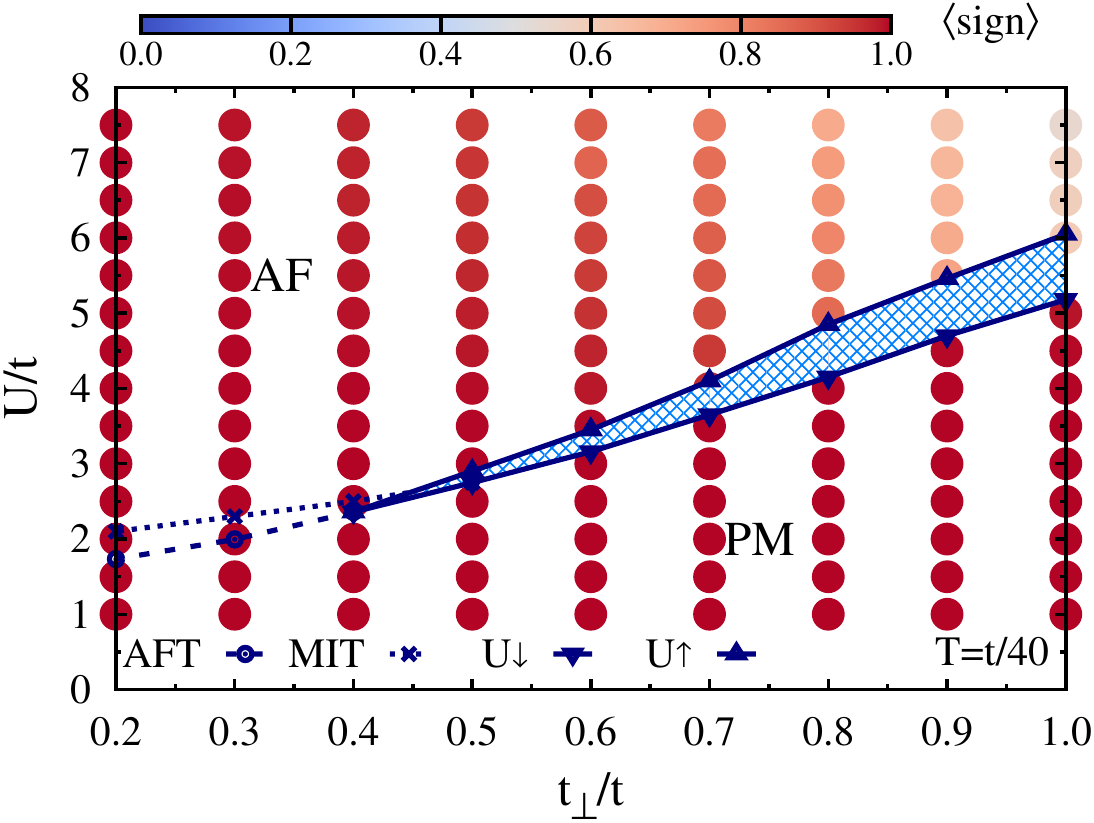}
\end{center}
\caption
{Same as in Fig.~\ref{PD_cdmft} but with color-coded circles displaying the behavior of the average sign in CT-QMC simulations 
within the $2\times 2$ CDMFT at $T=t/40$.}
\label{PD_sign}
\end{figure}

At half-filling the CT-QMC solver is sign free in the particle-hole symmetric case.
However, a finite value of the next-nearest-neighbor hopping $t'=-t_{\perp}/2$ used in our studies breaks the 
particle-hole symmetry and leads to a negative sign problem. 
The average sign in CT-QMC simulations within the $2\times 2$ CDMFT framework across the phase diagram at  
$T=t/40$ is shown in Fig.~\ref{PD_sign}. As is apparent, the sign problem is most severe in the isotropic 2D case. 
However, it becomes milder upon increasing the lattice anisotropy such that at $t_{\perp}/t=0.3$ we did not observe 
it around the magnetic transition point down to our lowest $T=t/100$. 
In this case, the limiting factor comes from the $(\beta N_c)^3$ scaling of the CT-QMC cluster solver.

\section{\label{app:hysteretic}  Evidence of hysteretic behavior around the first-order transition}

\begin{figure}[t!]
\begin{center}
\includegraphics[width=0.4\textwidth]{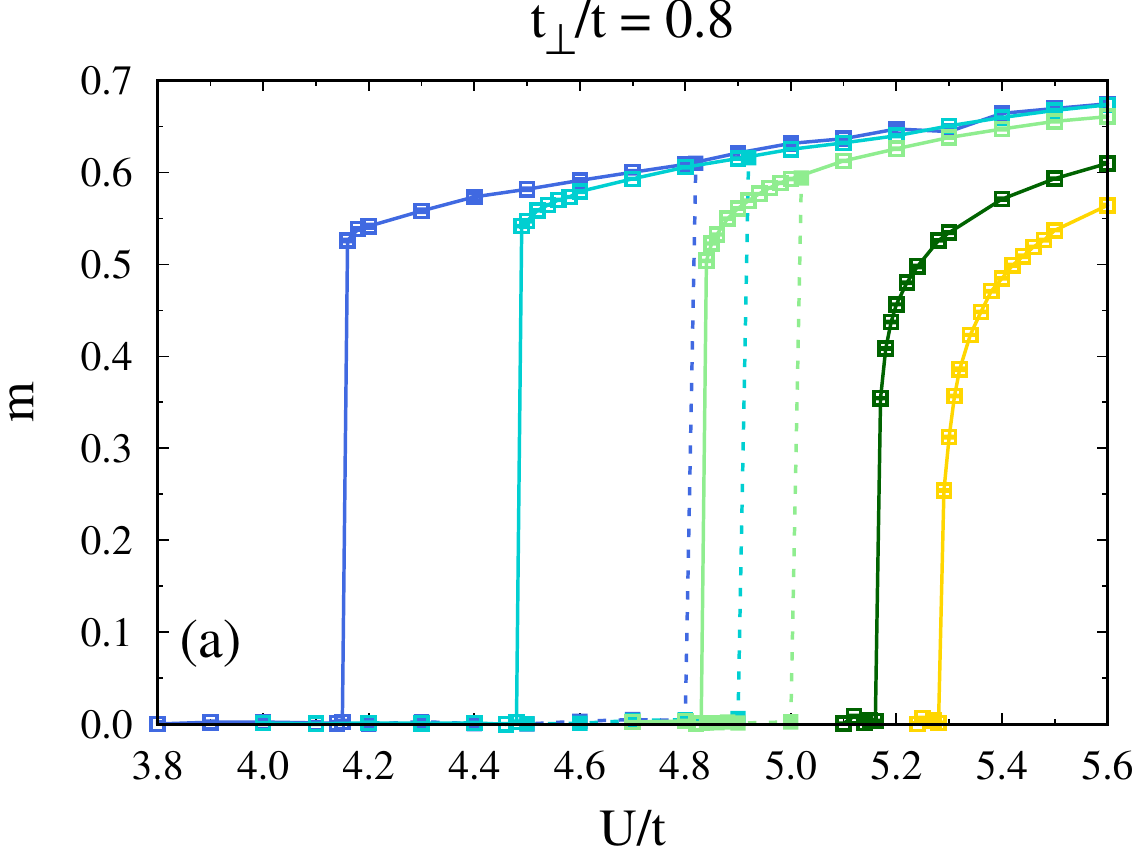}\\
\includegraphics[width=0.4\textwidth]{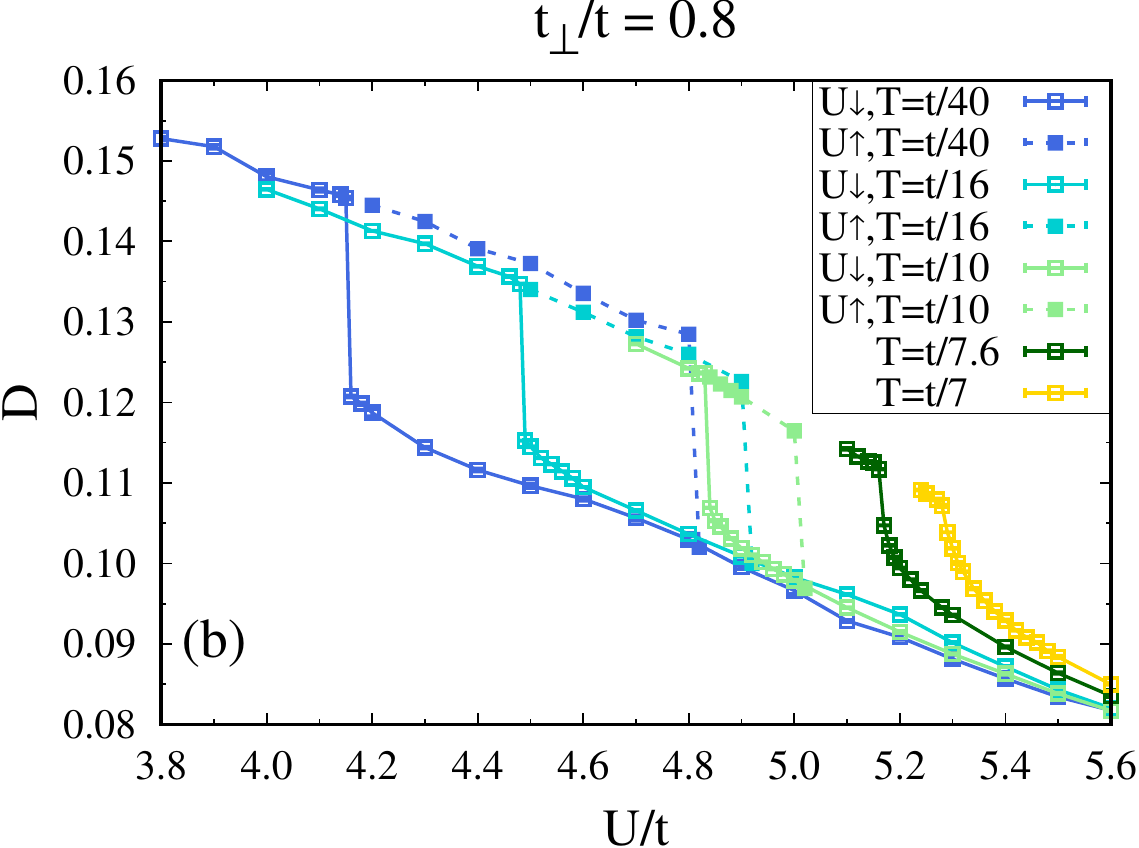}
\end{center}
\caption
{Staggered magnetization $m$ (a) and double occupancy $D$ (b) measured in decreasing ($U\hspace{-0.3em}\downarrow$, solid line) 
and increasing ($U\hspace{-0.3em}\uparrow$, dashed line) interaction sweeps at  various temperatures for $t_{\perp}/t=0.8$.  
}
\label{hyster_08}
\end{figure}

\begin{figure}[t!]
\begin{center}
\includegraphics[width=0.4\textwidth]{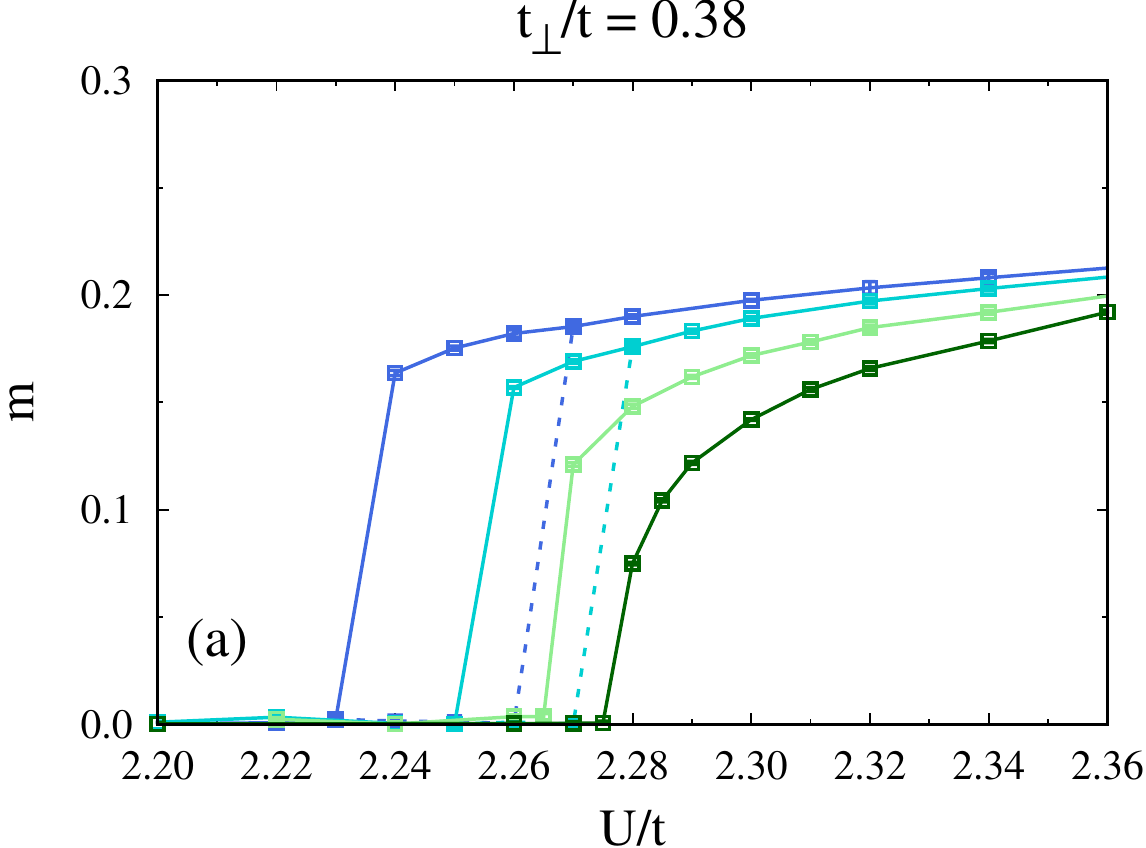}\\
\includegraphics[width=0.4\textwidth]{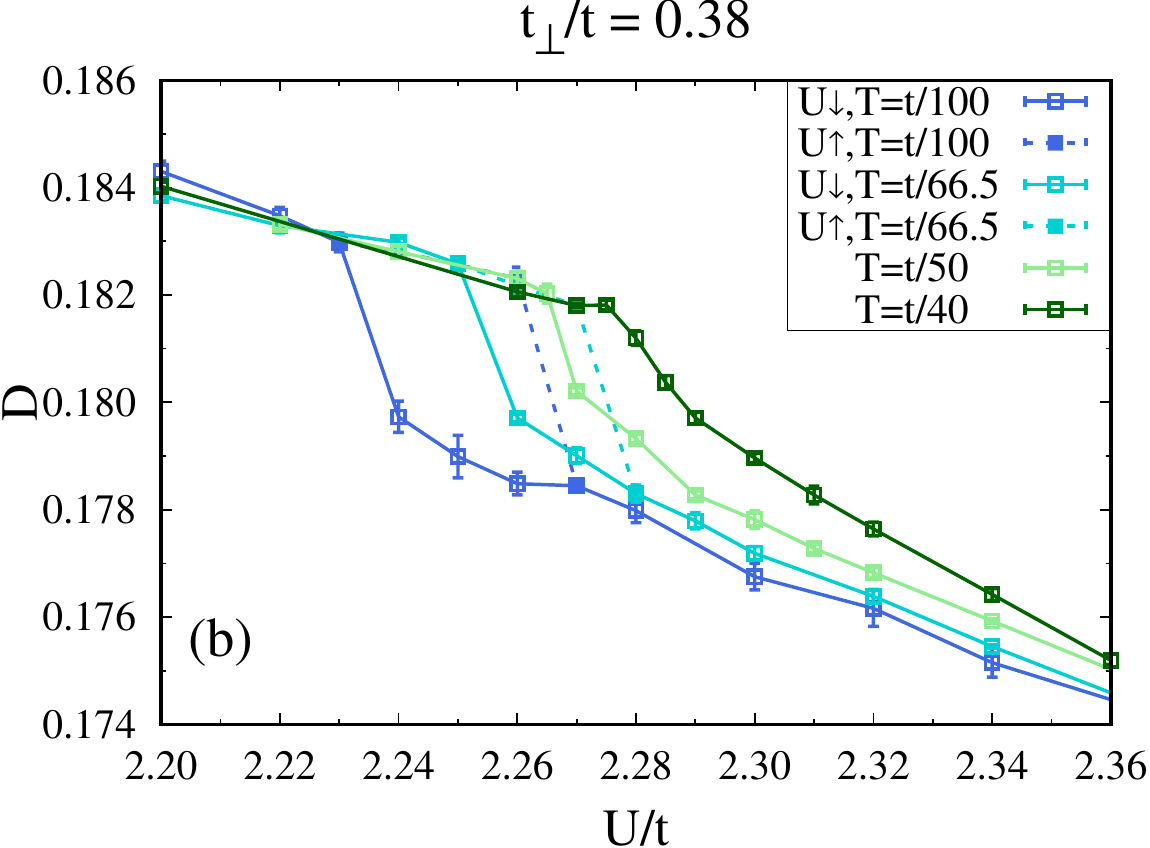}
\end{center}
\caption
{Same as in Fig.~\ref{hyster_08} but for $t_{\perp}/t=0.38$.
}
\label{hyster_038}
\end{figure}

In this Appendix, we provide numerical evidence for hysteretic behavior around the transition point in the AF magnetization 
$m$ and double occupancy $D$ measured in decreasing ($U\hspace{-0.3em}\downarrow$) and increasing ($U\hspace{-0.3em}\uparrow$)
interaction sweeps. As an example,  Fig.~\ref{hyster_08} shows the raw data measured at $t_{\perp}/t=0.8$ at various temperatures.  
In the $U\hspace{-0.3em}\downarrow$ sweep corresponding to the AF $\to$ PM transition (solid line in Fig.~\ref{hyster_08}), 
a converged AF solution at a given $U$ was used as an input for the next CDMFT simulation with a slightly smaller $U$. 
In the $U\hspace{-0.3em}\uparrow$ sweep corresponding to the PM $\to$ AF transition (dashed line in Fig.~\ref{hyster_08}), 
we initialized the CDMFT loop with a small staggered field; with increasing number of iterations, the solution evolved then
either into a PM or AF state. Repeating this procedure at gradually larger $U$ allowed us to find the second branch of 
the hysteresis loop below the critical end point $T_c$. 
For comparison, we also provide in Fig.~\ref{hyster_038}  the raw data for $t_{\perp}/t=0.38$. In this case, the magnetic transition 
at $T=t/40$ is continuous and one has to use lower temperatures $T\lesssim t/66.5$ to reveal a narrow hysteretic behavior.

\bibliographystyle{bibstyle}
\bibliography{marcin_refs}

\end{document}